\documentclass[fleqn,usenatbib]{mnras}
\usepackage[T1]{fontenc}
\usepackage{ae,aecompl}
\usepackage{color}
\usepackage{graphicx}
\usepackage{natbib}
\newcommand{\yhteys}[1]{{#1}}
\newcommand{\Data}[2]{{{\sc data}$_{\mathrm{{#1},{#2}}}$}}

\newcommand{\flip}{flip--flop}
\newcommand{\Inc}{incompatibility}
\newcommand{\cpsmodel}{CPS-model}
\newcommand{\cpsmethod}{CPS-method}
\newcommand{\twopmodel}{2P-model}
\newcommand{\twopmethod}{2P-method}
\newcommand{\onepmodel}{1P-model}
\newcommand{\onepmethod}{1P-method}
\newcommand{\kmethod}{K-method}
\newcommand{\wkmethod}{WK-method}
\newcommand{\sdr}{SDR}
\newcommand{\simethod}{SI-method}
\newcommand{\lcmethod}{LC-method}
\newcommand{\lsmethod}{LS-method}
\newcommand{\lsmodel}{LS-model}

\newcommand{\problemone}{{\color{black}   \it 1st question}}
\newcommand{\problemtwo}{{\color{black}   \it 2nd question}}
\newcommand{\problemthree}{{\color{black} \it 3rd question}}
\newcommand{\problemfour}{{\color{black}  \it 4th question}}
\newcommand{\problemfive}{{\color{black}  \it 5th question}}
\newcommand{\problemsix}{{\color{black}   \it 6th question}}
\newcommand{\JHLarg}{{JHL-ar\-gu\-ment}}
\newcommand{\JHLtext}{
{``The {\it observed} light curves of chromospherically active binary
and single stars are interference of two {\it real} 
constant period light curves of long-lived starspots.
These constant periods are the non-stationary active longitude
period $P_{\mathrm{act}}$ and the stationary rotation period
$P_{\mathrm{rot}} \approx P_{\mathrm{orb}}$.''}}
\newcommand{\questionone}{\problemone: 
Why have these two constant periods,
the non-stationary active longitude period $P_{\mathrm{act}}$ 
and 
the stationary rotation period $P_{\mathrm{rot}} \approx P_{\mathrm{orb}}$ 
of the
\JHLarg,
{\it not} been detected in the 
surface images?
}
\newcommand{\questiontwo}{\problemtwo: 
Why have these two constant periods,
the non-stationary active longitude period $P_{\mathrm{act}}$ 
and 
the stationary rotation period $P_{\mathrm{rot}} \approx P_{\mathrm{orb}}$ 
of
the \JHLarg, {\it not} 
been detected in the light curves?
Why are so {\it many different} 
$P_{\mathrm{rot}} \approx P_{\mathrm{phot}}$ 
periods {\it observed} 
in the light curves of the {\it same} star,
if the photometric 
data contains only these two constant periods?
}
\newcommand{\questionthree}{\problemthree:
Why do the surface images and 
the light curves give {\it different}
surface differential rotation estimates 
even for the {\it same} 
individual star?
}
\newcommand{\questionfour}{ \problemfour: 
\citet{Hac13} applied  the Kuiper method to
the $t_{\mathrm{min,1}}$ light curve minima  of FK Com.
Does this kind of an analysis give 
an {\it unambiguous} estimate for 
the non-stationary active longitude period
$P_{\mathrm{act}}$ of the \JHLarg?
}
\newcommand{\questionfive}{\problemfive: 
The Kuiper method detects the non-stationary active longitude
period $P_{\mathrm{act}}$
of the \JHLarg ~from 
the seasonal $t_{\mathrm{min,1}}$ light curve minima  
of chromospherically active binary 
stars \citep[][]{Jet17}
and single stars \citep[][]{Hac13}.
Why does this method {\it not} detect 
the stationary rotation period $P_{\mathrm{rot}}\approx P_{\mathrm{orb}}$
of the \JHLarg?
The long-term mean light curves of 
chromospherically active binary stars
in \citet{Jet17} follow the 
stationary rotation period $P_{\mathrm{rot}} \approx P_{\mathrm{orb}}$
of the \JHLarg.
Why do these long-term mean light curves {\it not} follow
the non-stationary active longitude period $P_{\mathrm{act}}$
of the \JHLarg?
}
\newcommand{\questionsix}{\problemsix:
What explains the observed {\it abrupt} \flip ~events in 
the chromospherically active stars,
if the long-lived starspots rotate with the two {\it regular 
constant} periods,
the non-stationary active longitude period $P_{\mathrm{act}}$ 
and 
the stationary rotation period
$P_{\mathrm{rot}} \approx P_{\mathrm{orb}}$?
}

\newcommand{\testone}{{\sc{test}}$_{a_1=a_2}$}
\newcommand{\testtwo}{{\sc{test}}$_{a_1<a_2}$}
\newcommand{\firstcase}{{{\sc \color{black} case}$_1$}}
\newcommand{\secondcase}{{{\sc \color{black} case}$_2$}}
\newcommand{\thirdcase}{{{\sc \color{black} case}$_3$}}

\newcommand{\migrone}{{{\sc migration}$_1$}}
\newcommand{\migrtwo}{{{\sc migration}$_2$}}
\newcommand{\migrthree}{{{\sc migration}$_3$}}

\newcommand{\paperi}{{\sc paper~i}}
\newcommand{\paperii}{{\sc paper~ii}}

\title[Real light curves of FK Com]
{Real light curves of FK Comae Berenices:  
Dead end }
\author[L. Jetsu]{
L. Jetsu\thanks{E-mail: lauri.jetsu@helsinki.fi} 
\\
$$Department of Physics, 
P.O. Box 64, FI-00014 University of Helsinki, 
Finland}
\date{Accepted XXX. Received YYY; in original form ZZZ}
\pubyear{2018}
\begin{document}
\label{firstpage}
\pagerange{\pageref{firstpage}--\pageref{lastpage}}
\maketitle
\begin{abstract}
Recently, we presented
a general model for the 
light curves
of chromospherically active stars,
where the observed light curve
is 
interference 
of two real constant period light curves 
of long-lived starspots. 
In this first paper, we make
six specific questions which undermine
this argument, because
it contradicts the current widely 
held views about the stellar 
surface differential
rotation and the starspots. 
Our aim is to answer 
these six questions.
We present evidence
that the long-lived starspots
of our general model
have already been detected in
the earlier surface imaging studies. 
The Lomb-Scargle power spectrum method analysis
of the real and the simulated data of FK Com reveals
that this 
method fails to detect the two real constant period light
curves of our general model.
{\it If} our model is valid,
this method gives
incompatible period, amplitude 
and minimum
epoch estimates 
telling nothing about the real periods, 
the real amplitudes
and the real minimum epochs of the two 
real light curves. 
This would mean that
all earlier one-dimensional period 
analyses
of the light curves of chromospherically active stars
have given spurious results
which have been widely and
uncritically accepted 
since the discovery of the starspots 
in the year 1947.
However, we arrive at a dead end,
because we can not solve the real light curves of FK Com.
In our second paper,
we solve these real light curves with
a new two-dimensional period finding method,
prove the validity of our general model,
and answer {\it all} 
six questions made in this first paper.
\end{abstract}


\section{Introduction}
\label{oneintro}

The ancient Egyptian papyrus Cairo 86637, 
which contains the calendar of lucky and unlucky days,
is the oldest preserved historical 
document of the discovery of a variable star, Algol
\citep[][]{Por08,Jet13,Jet15,Por18}.
The Sun is also a variable star.
The luminosity changes of Algol can be observed with
naked eye, but those of the Sun can be 
reliably confirmed only with satellite observations
\citep{Wil91,Rad18}.
The oldest preserved drawing of a sunspot was made by 
John of Worcester in the year 1128 \citep[][]{van96}.
\citet{Sch44} discovered the eleven years cycle in the number 
of sunspots.
The Zeeman effect caused by the solar magnetic field
was discovered by \citet{Hal08},
who also found out that this magnetic field 
is stronger in the sunspots.

\citet{Kro47} discovered the 
starspots in the light curves of 
the eclipsing binary AR Lacertae.
He observed short-term light curve
changes  ``within a few weeks to a few months''.
FK Comae Berenices (HD 117555, FK Com)
was among the first late--type stars where 
the starspots were also discovered 
\citep{Chu66}.
This chromosperically active single G4 giant \citep{Str09}  
is the prototype of a class of variable stars,
the FK~Com class, defined by \citet{Bop81}
as rapidly rotating single G--K giants.
Only a few stars belonging to this class have been found
\citep[][]{Puz14,How16,Puz17}. 
The members of this class may represent
recently coalesced W~UMa binaries \citep{Web76,Bop81,Egg89,Wel94}.

The starspots on the surface of FK Com seem to
concentrate
on two long--lived active longitudes separated by 180 degrees, 
and undergo abrupt shifts between these longitudes.
These shifts are called the \flip ~events
\citep{Jet91,Jet93}.
This phenomenon has also been observed
in the chromospherically active binaries
\citep[e.g.][$\sigma$ Gem]{Jet96}.
\citet{Hac13} applied the Kuiper method
(hereafter the \kmethod) to the seasonal light curve
minimum epochs of FK Com.
This analysis gave the active longitude period 
$P_{\mathrm{act}}=2.^{\mathrm{d}}401155 \pm 0.^{\mathrm{d}}000092$.
Numerous photometric studies of FK Com 
have been made \citep[e.g.][]{Kor02,Ola06,Pan07,Hac13}.
The starspot distribution of FK Com has also been
mapped with the surface imaging methods
\citep[e.g.][]{Kor99,Kor00,Kor04,Kor07,Kor09,Hac13,Vid15}.

\citet{Jet17} presented a general light curve
model for the Chromospherically Active Binary Stars 
(hereafter CABS).
They studied the long-term 
Mean Light Curve (hereafter MLC)
of fourteen CABSs as a function of orbital phase,
and also showed how their new model can be applied 
to the photometry of Chromospherically Active 
Single Star (hereafter CASS),
like FK Com.

\citet[][JHL = Jet\-su, Henry, Leh\-ti\-nen]{Jet17} made 
the following argument

\begin{description}

\item[-] \JHLarg: \JHLtext

\end{description}
At first sight, this \JHLarg ~would
seem to contradict the current overwhelming 
observational evidence for 
stellar Surface Differential Rotation (hereafter \sdr).
There are two widely used observational 
methods for measuring stellar SDR \citep[][Sect. 7]{Str09}.
In the first method,
the rotation periods $P_{\mathrm{rot}}$
of starspots at different latitudes
are measured from the Surface Images
\citep[hereafter the \simethod, e.g.][]{Pet04,Bar05,Col07,Str09,Bal16,Kov17}.
The second method measures the range
of the photometric rotation period $P_{\mathrm{phot}} \approx P_{\mathrm{rot}}$
changes in the Light Curves
\citep[hereafter the 
\lcmethod, e.g.][]{Hal90,Hal91A,Rei13A,Rei13,Rei15,Leh16,Dis16}.
There are numerous studies, where the \simethod ~and 
the \lcmethod ~give 
{\it different} \sdr ~estimates even 
for the {\it same} individual star.
For example, the \simethod ~study by \citet{Kor00} 
indicated ``solid body rotation'' in FK Com,
but photometric rotation
period changes of about 3.1\% were measured
with the \lcmethod ~by \citet[][$Z \approx 0.0308$]{Hac13}.

If the \JHLarg ~is true, then we must be able to answer
at least five of the six undermining questions made below.
The validity our argument does not
depend on the \problemthree.
On the contrary, the \JHLarg ~provides an
answer to this question.

In our Abstract, we refer
to these six specific questions:

\begin{description}

\item[-] \questionone 

\item[-] \questiontwo

\item[-] \questionthree

\item[-] \questionfour

\item[-] \questionfive

\item[-] \questionsix

\end{description}

\noindent
In this first paper, we can not
answer {\it all} these six questions.
In our next second paper, 
we present a new period analysis method 
based on the \JHLarg, apply this method to the
photometry of FK Com and  give compact answer to {\it all} 
the above six questions
\citep[][hereafter \paperii]{Jet18B}.

Our Appendix gives
all abbreviations used in this paper.


\begin{table}
\caption{Discarded data. 
Observing time (HJD), 
magnitude ($V$),
telescope (TEL),
segment (SEG) and
discarding criterion (C)}
\begin{center}
\renewcommand{\arraystretch}{0.92}
\begin{tabular}{ccccc}
\hline
HJD & $V$ & TEL & SEG & C \\
\hline
   2449876.7555 &    8.213 &    2 &   -- &    1 \\ 
   2449878.6786 &    8.109 &    2 &   -- &    1 \\ 
   2449882.6775 &    8.192 &    2 &   -- &    1 \\ 
   2449883.6783 &    8.158 &    2 &   -- &    1 \\ 
   2449883.7229 &    8.155 &    2 &   -- &    1 \\ 
   2449886.7003 &    8.309 &    2 &   -- &    1 \\ 
   2449886.7292 &    8.304 &    2 &   -- &    1 \\ 
   2449887.6784 &    8.180 &    2 &   -- &    1 \\ 
   2449888.6833 &    8.203 &    2 &   -- &    1 \\ 
   2449891.6809 &    8.346 &    2 &   -- &    1 \\ 
   2449893.6815 &    8.260 &    2 &   -- &    1 \\ 
   2449894.6718 &    8.250 &    2 &   -- &    1 \\ 
   2449896.6696 &    8.330 &    2 &   -- &    1 \\ 
   2449898.6701 &    8.308 &    2 &   -- &    1 \\ 
   2449900.6693 &    8.217 &    2 &   -- &    1 \\ 
   2449901.6691 &    8.289 &    2 &   -- &    1 \\ 
   2449902.6708 &    8.139 &    2 &   -- &    1 \\ 
   2449903.6705 &    8.318 &    2 &   -- &    1 \\ 
   2449904.6692 &    8.167 &    2 &   -- &    1 \\ 
   2449905.6591 &    8.210 &    2 &   -- &    1 \\ 
   2449906.6593 &    8.217 &    2 &   -- &    1 \\ 
   2449907.6589 &    8.161 &    2 &   -- &    1 \\ 
   2449908.6573 &    8.329 &    2 &   -- &    1 \\ 
   2450130.0134 &    8.094 &    2 &    1 &    5 \\ 
   2450504.8958 &    8.799 &    2 &    2 &    4 \\ 
   2450866.8368 &    8.154 &    1 &    1 &    3 \\ 
   2450934.6557 &    8.264 &    1 &    1 &    4 \\ 
   2450955.7222 &    8.144 &    1 &    1 &    3 \\ 
   2450974.6695 &    8.284 &    1 &    1 &    4 \\ 
   2450979.6702 &    8.322 &    1 &    1 &    4 \\ 
   2450986.7320 &    8.315 &    1 &    1 &    4 \\ 
   2451278.7070 &    8.104 &    1 &    2 &    3 \\ 
   2451278.7273 &    8.120 &    1 &    2 &    3 \\ 
   2451593.8157 &    8.217 &    2 &    5 &    5 \\ 
   2451706.7363 &    8.353 &    1 &    3 &    2 \\ 
   2451731.6859 &    8.195 &    2 &    5 &    2 \\ 
   2452251.0008 &    8.383 &    2 &    7 &    4 \\ 
   2452258.0132 &    8.386 &    2 &    7 &    4 \\ 
   2452348.8437 &    8.190 &    2 &    7 &    5 \\ 
   2452460.7124 &    8.324 &    2 &    7 &    2 \\ 
   2452461.6795 &    8.195 &    2 &    7 &    2 \\ 
   2452790.7993 &    8.248 &    1 &    6 &    5 \\ 
   2452807.6800 &    8.104 &    2 &    8 &    3 \\ 
   2452815.7371 &    8.288 &    1 &    6 &    2 \\ 
   2452824.7039 &    8.179 &    1 &    6 &    2 \\ 
   2452826.6991 &    8.171 &    1 &    6 &    2 \\ 
   2452828.6936 &    8.249 &    1 &    6 &    2 \\ 
   2453046.8944 &    8.013 &    2 &    9 &    5 \\ 
   2453070.8216 &    8.298 &    1 &    7 &    5 \\ 
   2453074.7515 &    7.999 &    2 &    9 &    3 \\ 
   2453139.6703 &    8.088 &    1 &    7 &    2 \\ 
   2453140.7700 &    8.379 &    2 &    9 &    4 \\ 
   2453144.6940 &    8.135 &    1 &    7 &    2 \\ 
   2453182.6611 &    8.060 &    1 &    7 &    2 \\ 
   2453184.6607 &    8.232 &    1 &    7 &    2 \\ 
   2453185.7171 &    8.155 &    1 &    7 &    2 \\ 
   2453709.0464 &    8.256 &    1 &    9 &    2 \\ 
   2453711.0410 &    8.346 &    1 &    9 &    2 \\ 
   2453711.0474 &    8.345 &    1 &    9 &    2 \\ 
   2453922.6616 &    8.243 &    1 &    9 &    2 \\ 
   2453922.6978 &    8.254 &    1 &    9 &    2 \\ 
   2454540.7813 &    8.087 &    1 &   11 &    3 \\ 
   2454807.0399 &    8.220 &    1 &   12 &    2 \\ 
   2454811.0305 &    8.260 &    1 &   12 &    2 \\ 
   2454822.0034 &    8.205 &    1 &   12 &    2 \\ 
   2454822.0349 &    8.254 &    1 &   12 &    2 \\ 
   2454999.7593 &    8.341 &    1 &   12 &    2 \\ 
   2455004.7444 &    8.298 &    1 &   12 &    2 \\ 
\hline
\end{tabular}
\renewcommand{\arraystretch}{1.00}
\end{center}
\label{tablerejected}
\end{table}

\section{Data}
\label{data}

We analyse the standard Johnson $V$ photometry of FK Com
published by \citet[][]{Hac13}.
These observations were made with 
the ``T7'' and ``Ph10'' telescopes.
We denote these separate samples of observations 
with TEL=1 (``T7'') and TEL=2 (``Ph10'').
\cite{Hac13} stored them as two separate files 
into the CDS database.
We analyse these two files separately,
to avoid bias, as \citet{Hac13} also did.
The accuracy of this photometry is between $0.^{\mathrm{m}}004$
and $0.^{\mathrm{m}}008$ in good photometric nights
\citep{Hen95A,Str97}.
All observations where the standard deviation of three
measurements exceeds $0.^{\mathrm{m}}020$ are automatically discarded.

We discard $n=68$ observations (Table \ref{tablerejected}).
The number of remaining analysed observations is $n=3807$.

The {\it isolated} first segment of TEL=2 data in \citet{Hac13} 
contains too few observations for reliable modelling,
and it is discarded 
(Table \ref{tablerejected}: C=1, $n=23$).

We divide the remaining data into seasonal segments 
(Table \ref{tabledata}: SEG).
These segment numbers are also used for
the rest of the discarded data in Table \ref{tablerejected}.

Some {\it isolated} observations are discarded in the beginning 
or in the end of a few segments 
(Table \ref{tablerejected}: C=2, $n=24$).
These isolated observations would
mislead the second order polynomial fits which are
used to eliminate the changes of the mean brightness
within segments (Sect. \ref{models}: Eq. \ref{meanoff}).
The fraction of all {\it isolated} discarded observations, $n=47$,
is 1.2\% in all data.

We use a sliding window to identify
the first group of {\it outliers}.
The mean ($m$) and the standard deviation ($s$) is
computed for all observations 
that are within $t \pm 30$ days 
of each individual observation $V(t)$. 
We discard those individual $V(t)$ 
observations that are
below $m-2.5s$ or above $m+2.5s$.
Some observations below the $m-2.5s$ limit
may represent photometric flares 
(Table \ref{tablerejected}: C=3, $n=7$).
These events have been observed 
earlier in FK Com
\citep{Jet93}.
All observations above the $m+2.5s$ limit
are very probably erroneous
(Table \ref{tablerejected}: C=4, $n=8$).

If two observations made during the {\it same} night
with the {\it same} telescope deviate more than 
$0.^{\mathrm{m}}06$, we discard one of them as an 
{\it outlier}. 
This second group of outliers
contains only $n=6$ observations
(Table \ref{tablerejected}: C=5).
The fraction of all {\it outliers}, $n=21$,
is 0.5\% in all data.

All the above mentioned outliers are identified
{\it before} our analysis. 
This analysis
will reveal that there are many other outliers
(\yhteys{\paperii: Sects. 3 and 4}).
However, we do not reject those outliers,
because they are identified only {\it after} the analysis.

\begin{table}
\caption{Standard Johnson $V$ photometry of FK Com. 
Telescope (TEL), 
segments (SEG),
first and last observing time ($t_1$ and $t_n$),
time span and number of observations ($\Delta T$ and $n$)}
\begin{center}
\begin{tabular}{cccccc}
\hline
TEL              &
SEG              & $t_1$          & $t_n$        & $\Delta T$ & $n$ \\
                 &
                 & HJD            & HJD          &  d & \\
\hline
   1 &    1 & 2450850.875 & 2450997.687 &    146.8 &    221 \\ 
   1 &    2 & 2451153.047 & 2451360.673 &    207.6 &    378 \\ 
   1 &    3 & 2451534.046 & 2451694.803 &    160.8 &     67 \\ 
   1 &    4 & 2451883.044 & 2452078.743 &    195.7 &     88 \\ 
   1 &    5 & 2452248.044 & 2452461.698 &    213.7 &    220 \\ 
   1 &    6 & 2452614.048 & 2452790.765 &    176.7 &    170 \\ 
   1 &    7 & 2452978.048 & 2453144.694 &    166.6 &    127 \\ 
   1 &    8 & 2453343.046 & 2453565.681 &    222.6 &    192 \\ 
   1 &    9 & 2453721.051 & 2453907.661 &    186.6 &    189 \\ 
   1 &   10 & 2454075.045 & 2454285.686 &    210.6 &    181 \\ 
   1 &   11 & 2454440.049 & 2454643.730 &    203.7 &    187 \\ 
   1 &   12 & 2454807.040 & 2455004.744 &    197.7 &    129 \\ 
   1 &   13 & 2455172.044 & 2455297.831 &    125.8 &    111 \\ 
   2 &    1 & 2450085.055 & 2450265.674 &    180.6 &    111 \\ 
   2 &    2 & 2450412.038 & 2450636.678 &    224.6 &    209 \\ 
   2 &    3 & 2450778.042 & 2450997.706 &    219.7 &    205 \\ 
   2 &    4 & 2451144.039 & 2451362.728 &    218.7 &    179 \\ 
   2 &    5 & 2451508.043 & 2451712.687 &    204.6 &    157 \\ 
   2 &    6 & 2451873.042 & 2452089.727 &    216.7 &    193 \\ 
   2 &    7 & 2452242.033 & 2452448.752 &    206.7 &    188 \\ 
   2 &    8 & 2452613.009 & 2452826.679 &    213.7 &    159 \\ 
   2 &    9 & 2452972.029 & 2453194.682 &    222.7 &    154 \\ 
\hline
\end{tabular}
\end{center}
\label{tabledata}
\end{table}


\cite{Hac13} applied the Continuous Period Search 
method (hereafter the \cpsmethod)
to this photometry. 
We refer to the model of this method as the 
CPS-model \citep[][their Eq. 3]{Leh12}.
\cite{Hac13} modelled 1464 subsets of photometry,
but they did not publish their numerical values.
Their light curve periods, amplitudes, primary minimum epochs
and secondaryminimum epochs are now published in electronic form
\footnote{The full
Table \ref{electric} is only available via anonymous ftp  to
cdsarc.u-strasbg.fr (130.79.128.5) or via
http://cdsarc.u-strasbg.fr/viz-bin/qcat?J/...}
(Table \ref{electric}).

\begin{table*}
\caption{
CPS-method results from \citet{Hac13}.
The first column gives the telescope (TEL).
The remaining columns give the same subset parameters as
described in the Appendix of 
\citet[][their Table A.1]{Leh11}:
first observing time $(t_1)$,
last observing time $(t_n)$,
mean observing time $(\tau)$,
statistically independent estimates (IND: 1=Yes, 0=No),
number of observations $(n)$,
period $(P \pm \sigma_P)$,
peak to peak light curve amplitude $(A \pm \sigma_A)$,
epochs of primary and secondary minima 
$(t_{\mathrm{min,1}} \pm \sigma_{\mathrm{t_{min,1}}} ,
t_{\mathrm{min,2}} \pm \sigma_{\mathrm{t_{min,2}}})$.
The dummy value ``-1.000'' denotes the
cases where no estimate was obtained.
The units of Heliocentric Julian Days are HJD-2~400~000. 
Note that we show only the ten first lines of this table.
The full table is published only in electronic form.
It contains 1464 lines (TEL=1: 760 lines) and (TEL=2: 704 lines).}
\begin{center}
\addtolength{\tabcolsep}{-0.05cm} 
\begin{tabular}{cccccccccccccc}
\hline
TEL & $t_1$ & $t_n$ & $\tau$ & IND & $n$ & 
$P$ & $\sigma_P$ &
$A$ & $\sigma_A$ &
$t_{\mathrm{min,1}}$ & $\sigma_{\mathrm{t_{min,1}}}$ &
$t_{\mathrm{min,2}}$ & $\sigma_{\mathrm{t_{min,2}}}$ \\
    & HJD & HJD & HJD &  &  & d & d & mag & mag 
    & HJD & d 
    & HJD & d \\
\hline
  1 &   50850.875 &   50874.887 &   50863.555 &  1 &  29 & 2.4129 & 0.0036 &  0.059 &  0.006 &   50851.594 &   0.032 &      -1.000 &  -1.000 \\ 
  1 &   50851.871 &   50875.883 &   50865.793 &  0 &  28 & 2.4154 & 0.0029 &  0.059 &  0.004 &   50854.043 &   0.034 &      -1.000 &  -1.000 \\ 
  1 &   50852.867 &   50876.883 &   50867.895 &  0 &  29 & 2.4141 & 0.0036 &  0.051 &  0.004 &   50853.980 &   0.031 &      -1.000 &  -1.000 \\ 
  1 &   50855.867 &   50878.809 &   50870.176 &  0 &  31 & 2.4196 & 0.0033 &  0.050 &  0.003 &   50856.395 &   0.027 &      -1.000 &  -1.000 \\ 
  1 &   50856.863 &   50880.867 &   50872.070 &  0 &  32 & 2.4155 & 0.0034 &  0.047 &  0.003 &   50858.812 &   0.029 &      -1.000 &  -1.000 \\ 
  1 &   50861.852 &   50885.855 &   50876.176 &  0 &  39 & 2.3930 & 0.0037 &  0.044 &  0.002 &   50863.754 &   0.029 &      -1.000 &  -1.000 \\ 
  1 &   50867.836 &   50891.840 &   50878.883 &  0 &  43 & 2.3923 & 0.0078 &  0.042 &  0.003 &   50868.613 &   0.042 &   50870.176 &   0.060 \\ 
  1 &   50868.828 &   50892.840 &   50880.621 &  0 &  43 & 2.3852 & 0.0051 &  0.041 &  0.003 &   50871.016 &   0.035 &   50870.215 &   0.057 \\ 
  1 &   50872.820 &   50896.828 &   50884.023 &  0 &  49 & 2.3699 & 0.0059 &  0.040 &  0.004 &   50875.008 &   0.041 &   50873.477 &   0.036 \\ 
  1 &   50873.820 &   50897.832 &   50885.551 &  0 &  49 & 2.3732 & 0.0061 &  0.039 &  0.004 &   50875.004 &   0.051 &   50875.844 &   0.037 \\ 
\hline
\end{tabular}
\addtolength{\tabcolsep}{+0.05cm} 
\end{center}
\label{electric}
\end{table*}

\section{Models}
\label{models}

Our notation for a $V(t)$ magnitude observation made at an observing time
$t=t_i$ is $y_i=y(t_i)$. 
The time span of these $n$ observations is $\Delta T=t_n-t_1.$ 
We fit a second order polynomial $h(t)$ to the $y_i$ data 
of each individual segment. 
This polynomial is used to eliminate the seasonal
changes of the mean brightness of FK Com.
The analysed data are
\begin{eqnarray}
y_i^{,}=y(t_i)-h(t_i)=y_i-h_i.
\label{meanoff}
\end{eqnarray}
\noindent
We use this second order polynomial $h(t)$ in all segments, because
higher orders tend to fluctuate at the ends of some segments. 
It would not be consistent or objective to vary the order of this
polynomial in different segments. 
However, we will mention if this second
order polynomial can, or can not, reproduce the
changes of the mean brightness in individual segments
(\yhteys{\paperii: Sects. 3 and 4}).

The abbreviation \Data{x}{y} is hereafter used
for the TEL=x observations in segment SEG=y.
Our notations for the mean and the standard deviation
of $y_i^{,}$ are $m_{y^{'}}$ and $\sigma_{y^{'}}$.
We model each segment of 
the $y_i^{,}$ data with the two and one period models.

\subsection{Two period model (\twopmodel) }
\label{secttwop}

The complex \twopmodel ~has two parts. The first part is
\begin{eqnarray}
g_1(t)\!=\!g_1(t,\bar{\beta}_1)\!=\!
\sum_{k=1}^{K_1} 
B_k \cos (k 2 \pi f_1 t)
+
C_k \sin (k 2 \pi f_1 t),
\label{gone}
\end{eqnarray}
\noindent
where  $K_1=2$. 
Higher $K_1$ values are unnecessary,
because the detection of three 
minima in the photometric light curve is practically
impossible \citep[e.g.][their Sect. 3.2]{Leh11}.
The five free parameters are 
the amplitudes $B_1, B_2, C_1$ and $C_2$,
and the frequency $f_1$.
The frequency units are 
$[f]={\mathrm{d}}^{-1}$.
The vector of free parameters is $\bar{\beta}_1=[B_1,B_2,C_1,C_2,f_1]$.
In every segment,
we determine the following parameters of the $g_1(t)$ function
\begin{description}
\item $P_1=$ period $=f_1^{-1}$
\item $A_1=$ peak to peak amplitude
\item $t_{\mathrm{g1,min,1}}=$ epoch of primary (i.e. deeper) minimum
\item $t_{\mathrm{g1,min,2}}=$ epoch of secondary minimum (if present)
\end{description}
The units are
$[P_1]={\mathrm{d}}$,
$[A_1]={\mathrm{mag}}$ and
$[t_{\mathrm{g1,min,1}}]=$
$[t_{\mathrm{g1,min,2}}]=$ ${\mathrm{HJD}}-2~400~000$.
We use the epoch of the first primary and secondary minimum
within each individual segment.
The phases of $g_1(t)$ are computed from 
\begin{eqnarray}
\phi_1={\mathrm{FRAC}}[(t-t_0)/P_1],
\label{phaseone}
\end{eqnarray}
\noindent
where ${\mathrm{FRAC}}[x]$ removes the integer part of its argument $x$,
and $t_0=t_1=$ the first observing time of the segment data.

The second part of the \twopmodel ~is 
\begin{eqnarray}
g_2(t)\!=\!g_2(t,\bar{\beta}_2)
\!=\!
\sum_{k=1}^{K_2} 
D_k \cos (k 2 \pi f_2 t)
+
E_k \sin (k 2 \pi f_2 t),
\label{gtwo}
\end{eqnarray}
\noindent
where $K_2=2$ and $\bar{\beta}_2$ $=[D_1,D_2,E_1,E_2,f_2]$.
The following parameters 
\begin{description}
\item $P_2=$ period $=f_2^{-1}$
\item $A_2=$ peak to peak amplitude
\item $t_{\mathrm{g2,min,1}}=$ epoch of primary minimum
\item $t_{\mathrm{g2,min,2}}=$ epoch of secondary minimum (if present)
\end{description}
of the $g_2(t)$ function
are determined for every segment.
The units are the same as 
those used for the $g_1(t)$ function.
We compute the phases of $g_2(t)$ from 
\begin{eqnarray}
\phi_2={\mathrm{FRAC}}[(t-t_0)/P_2],
\label{phasetwo}
\end{eqnarray}
\noindent
where $t_0=t_1$, as in Eq. \ref{phaseone}.

These two parts are combined in the complex \twopmodel
\begin{eqnarray}
g_{\mathrm{C}}(t)\!=\!g_{\mathrm{C}}(t,\bar{\beta}_{\mathrm{C}})\!=\!g_1(t,\bar{\beta}_1)+g_2(t,\bar{\beta}_2).
\label{gtwomodel}
\end{eqnarray}
\noindent
This nonlinear model has $p_{\mathrm{\mathrm{C}}}=10$ free parameters
$\bar{\beta}_{\mathrm{C}}=[\bar{\beta}_1,\bar{\beta}_2]$.
Note that the functions $g_1(t)$ and $g_2(t)$ have unique phases
(Eqs. \ref{phaseone} and \ref{phasetwo}),
but no such phases exist for the $g_{\mathrm{C}}(t)$ function
which is constantly changing in time.

\subsection{One period model (\onepmodel) }
\label{sectonep}

The simple \onepmodel ~is 
\begin{eqnarray}
g_{\mathrm{S}}(t)\!=\!g_{\mathrm{S}}(t,\bar{\beta}_{\mathrm{S}})\!=\!
\sum_{k=1}^{K_3} 
F_k \cos (k 2 \pi f_3 t)
+
G_k \sin (k 2 \pi f_3 t),
\label{gonemodel}
\end{eqnarray}
\noindent
where $K_3=2$ and $\bar{\beta}_{\mathrm{S}}=[F_1,F_2,G_1,G_2,f_3]$.
There are $p_{\mathrm{S}}=5$ 
free parameters in this nonlinear model.

We determine the following parameters 
\begin{description}
\item $P_3=$ period $=f_3^{-1}$
\item $A_3=$ peak to peak amplitude
\item $t_{\mathrm{S,min,1}}=$ epoch of primary minimum
\item $t_{\mathrm{S,min,2}}=$ epoch of secondary minimum (if present)
\end{description}
of the $g_{\mathrm{S}}(t)$ function.
The units are the same as 
those of the $g_1(t)$ and $g_2(t)$ functions.
The phases of $g_{\mathrm{S}}(t)$ are
\begin{eqnarray}
\phi_3={\mathrm{FRAC}}[(t-t_0)/P_3],
\label{phasethree}
\end{eqnarray}
\noindent
where $t_0=t_1$, as in Eq. \ref{phaseone}.

The \onepmethod ~is a ``one-dimensional'' period
finding method because it searches for only one 
periodicity in the data. 
We will consistently use the notations
$f_3$, $P_3$ and $A_3$ also for the frequencies,
the periods and the amplitudes of all other one-dimensional period
finding methods that are mentioned in this study.

\subsection{Nested models}
\label{nested}

The complex model of Eq. \ref{gtwomodel} and 
the simple  model of Eq. \ref{gonemodel} are nested.
They represent the same model if
\begin{description}

\item \firstcase: $f_1=f_2$ in Eqs. \ref{gone} and \ref{gtwo}
\item \secondcase: $B_1=B_2=C_1=C_2=0$ in Eq. \ref{gone}
\item \thirdcase: $D_1=D_2=E_1=E_2=0$ in Eq. \ref{gtwo}

\end{description}

\noindent
We denote the cases where the complex model approaches
the simple model with
\begin{eqnarray}
g_{\mathrm{C}} \rightarrow g_{\mathrm{S}}.
\label{breaksdown}
\end{eqnarray}
The \twopmodel ~''breaks down''
when this happens.
The alternatives when the \twopmodel ~and 
the \onepmodel ~are the same model occur
in these three cases
\begin{eqnarray}
 f_1-f_2 & \rightarrow & 0 \label{1case} \\ 
 A_1/A_2 & \rightarrow & 0 \label{2case} \\
 A_2/A_1 & \rightarrow & 0. \label{3case} 
\end{eqnarray}

\noindent
The residuals of the complex model
\begin{eqnarray}
\epsilon_i=y_i^,-g_{\mathrm{C}}(t_i)
\label{epsilonc} 
\end{eqnarray}
give the sum of squared residuals
$R_{\mathrm{C}}=\sum_{i=1}^n\epsilon_i^2.$
The simple model $g_{\mathrm{S}}(t)$ residuals
\begin{eqnarray} 
\epsilon_i=y_i^,-g_{\mathrm{S}}(t_i)
\label{epsilons} 
\end{eqnarray}
give $R_{\mathrm{S}}=\sum_{i=1}^n\epsilon_i^2$.
We compute the test statistic 
\begin{equation}
F=
\left(
{
{R_{\mathrm{S}}}
\over
{R_{\mathrm{C}}}
}
-1
\right)
\left(
{
{n-p_{\mathrm{C}}-1}
\over
{p_{\mathrm{C}}-p_{\mathrm{S}}}
}
\right).
\label{fvalue} 
\end{equation}
The null hypothesis is
\begin{description}

\item $H_{\mathrm{0}}$: {\it ``The complex model $g_{\mathrm{C}}(t)$ 
does not provide
a significantly better fit than the simple model $g_{\mathrm{S}}(t)$.'' }

\end{description}

\noindent
Under $H_0$, the test statistic of Eq. \ref{fvalue}
has an $F$ distribution with 
$(\nu_1,\nu_2)$ degrees of freedom,
where $\nu_1=p_{\mathrm{C}}-p_{\mathrm{S}}$ and $\nu_2=n-p_{\mathrm{C}}$
\citep{Dra98}. 
The probability that $F$ reaches or exceeds a fixed level $F_0$
is called the critical level $Q_{F} = P(F \ge F_0)$. 
We will reject $H_0$, if and only if
\begin{eqnarray}
Q_F < \gamma_F=0.001,
\label{oneortwo}
\end{eqnarray}
where $\gamma_F$ is a pre-assigned significance level.
It represents the probability of falsely
rejecting $H_0$ when it is in fact true.

\section{Earlier \twopmodel ~analysis  }
\label{earlier}

If the frequencies $f_1$ and $f_2$ in Eqs. \ref{gone} and \ref{gtwo} 
are unknown free parameters, the $g_{\mathrm{C}}(t)$
model of Eq. \ref{gtwomodel} is nonlinear
and there is no unique least squares fit solution.
This $g_{\mathrm{C}}(t)$ model becomes linear,
if these $f_1$ and $f_2$ frequencies 
are fixed to some known numerical constant values,
like $f_1=1/P_{\mathrm{act}}$ and $f_2=1/P_{\mathrm{orb}}$
in \citet{Jet17}.
In this case, the solutions for the eight 
remaining free parameters
of a least squares fit,
the amplitudes $B_1,B_2,C_1,C_2,D_1,D_2,E_1$ and $E_2$ of
the $g_1(t,\bar{\beta}_1)$ and $g_2(t,\bar{\beta}_2)$ functions, 
are unambiguous.

\cite{Jet17} studied the long-term 
photometry of 14 CABSs with
the \twopmodel.
They showed that the long-term MLC of all these CABSs
followed the orbital period  $P_{\mathrm{orb}}$
\citep[][Figs. 1-14: panels ``b-g'']{Jet17}.
These CABSs 
rotate synchronously, $P_{\mathrm{orb}} \approx P_{\mathrm{rot}}$.
This means that the starspot distribution causing
the MLC of these stars is stationary in the rotating reference frame.
It was also shown that the line connecting the
centres of the two binary components intersects
the longitudes of this stationary part.

\cite{Jet17} analysed
the epochs of the primary minima 
$t_{\mathrm{min,1}}$ of the seasonal
light curves of CABSs
with the Kuiper method.
This analysis revealed the presence of 
long-lived active longitudes rotating
with a constant period of $P_{\mathrm{act}}$.
Due to the inequality 
$P_{\mathrm{act}} \neq P_{\mathrm{orb}}$,
the seasonal light curve minima 
connected to the former period migrate in 
the rotating reference frame, while the minima 
connected to the
latter period do not \citep[][Figs. 15-27: 
tilted non-stationary
and horizontal stationary lines 
in panels ``a'']{Jet17}.

Two segments of the photometry
of RS CVn binary $\sigma$ Gem 
were modelled by using 
$f_1=1/P_{\mathrm{act}}$ and $f_2=1/P_{\mathrm{orb}}$
in Eqs. \ref{gone} and \ref{gtwo}
\citep[][Figs. 30 and 31]{Jet17}.
It was argued that {\it all} observed 
light curves of CABSs
may be interference of two real light curves with 
periods of  
$P_{\mathrm{act}}$ (non-stationary in rotating reference frame)
and 
$P_{\mathrm{orb}} \approx P_{\mathrm{rot}}$ 
(stationary  in rotating reference frame).

\citet[][Fig. 32]{Jet17} also modelled 
one segment of FK Com data 
with the \twopmodel.
They first fixed $f_1=1/P_{\mathrm{act}}$ 
in Eq. \ref{gone}.
Their numerical value was
$P_{\mathrm{act}}=2.^{\mathrm{d}}401155 
\pm 0.^{\mathrm{d}}000092$
\citep{Hac13}.
Then, they tested $P_{\mathrm{rot}}$ values which were $\pm 15$\% at both
sides of $P_{\mathrm{act}}$. 
For each of these tested $f_2=1/P_{\mathrm{rot}}$ 
in Eq. \ref{gtwo},
they computed a test statistic 
$z_{\mathrm{test}}(f_2)$ \citep{Jet17}.
For their chosen tested 
$f_1=1/P_{\mathrm{act}}$ and $f_2=1/P_{\mathrm{rot}}$ values,
their test statistic $z_{\mathrm{test}}$ 
was equal to the test statistic 
that we will introduce later 
in \yhteys{\paperii ~(Sect. 2.1: Eq. 2)}.

\section{Measuring surface 
differential rotation (\sdr)}
\label{sdrsect}

\subsection{Solar \sdr}
\label{solarsdr}

\citet{How94} discussed the uncertainties in all measurements of
solar \sdr.
He emphasized that these measurements are
averages of {\it many} features over latitude,
and he then gave an example of the {\it real~} 
\sdr ~of {\it individual} features.
His ``sobering reminder'' illustrated the solar \sdr
~measurements of 36~708 sunspot groups  made at Mount Wilson between 
the years 1917 and 1985 \citep[][his Fig. 2]{How94}.
He emphasized that the latitude of any {\it individual}
sunspot group did not predict its rotation period.

\subsection{Stellar \sdr} 
\label{stellarsdr}

If the {\it individual} sunspot groups do not follow the 
law of solar \sdr ~(see Eq. \ref{rotationsun}),
then why should the {\it individual} starspots 
or starspot groups do so?
More than ten  {\it individual} \simethod ~surface temperature maps 
are available only for 
a few stars \citep[see][Table 2]{Str09}.
The most extensive \lcmethod ~studies 
are typically based on about one hundred
 {\it individual} statistically independent
light curve $P_{\mathrm{phot}}$ values 
\citep[e.g.][]{Jet99A,Leh16}.
Our intention is not to undermine 
the previous \simethod ~and \lcmethod ~studies,
but it is fair to remind about the uncertainties 
in these stellar \sdr ~measurements,
if they truly rely on the ``solar-stellar connection''.


\begin{table*}
\caption{Sample of thirteen CABSs in Jetsu et al. (2017). 
Rotation period $(P_{\mathrm{rot}}\approx P_{\mathrm{orb}})$,
active longitude rotation period $(P_{\mathrm{act}})$,
difference $(|P_{\mathrm{rot}}-P_{\mathrm{act}}|)$,
differential rotation coefficient (Eq. \ref{si_k}: $|k|$) 
and
angular velocity difference (Eq. \ref{si_omega}: $|\Delta \Omega|$).
Note that the parameters are given in the order of increasing 
$|\Delta \Omega|$.}
\label{falsediff}
\begin{tabular}{lccccc} 
\hline
Star & $P_{\mathrm{rot}} \approx P_{\mathrm{orb}}$ & $P_{\mathrm{act}}$ & $|P_{\mathrm{rot}}-P_{\mathrm{act}}|$ &
$|k|$ & $|\Delta \Omega|$ \\
     & $\mathrm{d}$    & $\mathrm{d}$    & $\mathrm{d}$    &
      &  ${\mathrm{rad~d}}^{-1}$                \\
\hline
              DM~UMa &
     $7.492\pm0.009$ &
   $7.4898\pm0.0008$ &
   $0.0022\pm0.0090$ &
   $0.0003\pm0.0012$ &
   $0.0002\pm0.0010$ \\
              HK~Lac &
  $24.4284\pm0.0005$ &
      $24.40\pm0.01$ &
     $0.028\pm0.010$ &
 $0.00116\pm0.00041$ &
 $0.00030\pm0.00011$ \\
           V1149~Ori &
$53.57465\pm0.00072$ &
      $53.14\pm0.06$ &
     $0.435\pm0.060$ &
   $0.0081\pm0.0011$ &
 $0.00096\pm0.00013$ \\
              EL~Eri &
    $48.263\pm0.206$ &
      $47.69\pm0.02$ &
       $0.57\pm0.21$ &
   $0.0119\pm0.0043$ &
 $0.00156\pm0.00056$ \\
              BM~CVn &
  $20.6252\pm0.0018$ &
    $20.513\pm0.006$ &
   $0.1122\pm0.0063$ &
 $0.00545\pm0.00031$ &
$0.001666\pm0.000093$ \\
              II~Peg &
$6.724333\pm0.000010$ &
   $6.7119\pm0.0007$ &
 $0.01243\pm0.00070$ &
 $0.00185\pm0.00010$ &
$0.001731\pm0.000098$ \\
        $\sigma$~Gem &
$19.604471\pm0.000022$ &
    $19.497\pm0.005$ &
   $0.1075\pm0.0050$ &
 $0.00550\pm0.00026$ &
$0.001767\pm0.000083$ \\
              HU~Vir &
$10.387522\pm0.000031$ &
    $10.419\pm0.001$ &
   $0.0315\pm0.0010$ &
$0.003026\pm0.000096$ &
$0.001827\pm0.000058$ \\
              XX~Tri &
$23.96924\pm0.00092$ &
      $23.77\pm0.01$ &
     $0.199\pm0.010$ &
 $0.00835\pm0.00042$ &
 $0.00220\pm0.00011$ \\
           V1762~Cyg &
$28.58973\pm0.00002$ &
      $28.17\pm0.02$ &
     $0.420\pm0.020$ &
 $0.01479\pm0.00071$ &
 $0.00327\pm0.00016$ \\
              FG~UMa &
$21.35957\pm0.00040$ &
      $21.12\pm0.01$ &
     $0.240\pm0.010$ &
 $0.01128\pm0.00047$ &
 $0.00334\pm0.00014$ \\
              EI~Eri &
$1.947227\pm0.000008$ &
   $1.9545\pm0.0008$ &
 $0.00727\pm0.00080$ &
 $0.00373\pm0.00041$ &
   $0.0120\pm0.0013$ \\
            V711~Tau &
 $2.83774\pm0.00001$ &
   $2.8924\pm0.0002$ &
 $0.05466\pm0.00020$ &
$0.019078\pm0.000069$ &
 $0.04184\pm0.00015$ \\
\hline
\end{tabular}
\end{table*}

\section{\simethod }
\label{simethodsect}

The compact answer to the \problemone ~in 
\yhteys{\paperii ~(Sect. 12)}
is based on results presented in the next
Sects. \ref{SIparameters}-\ref{SIparticular}.

\subsection{\simethod ~parameters}
\label{SIparameters}

In the \simethod,
the aim is to determine the rotation periods of 
recognizable starspot distribution patterns at different latitudes.
There are numerous different approaches
\citep[see][]{Str09,Str11,Koc16}.
For example, the rotation periods can be estimated directly
from the surface images \citep[e.g.][]{Bar05,Bal16},
or simultaneous photometry can be used as an additional
inversion constraint, or the latitudes of starspots 
in the surface images can be compared to the simultaneous 
light curves \citep[e.g.][]{Ber98B,Kor00,Hac04,Kor07,Col15}.

The \simethod ~inversion represents an ``ill posed problem'',
because an infinite number of different solutions can fit 
the spectroscopic observations.
Several stellar physical parameters are fixed {\it before} these inversions,
e.g. inclination, rotation period and local spectral line profiles.
This introduces additional uncertainty to the \simethod ~results
\citep[see][his Sect 9.2.3.]{Koc16}.
One of the crucial 
fixed parameters in these inversions
is the stellar rotation period $P_{\mathrm{rot}}$ value. 
It is fixed to
some constant numerical value, 
like $P_{\mathrm{phot}}$ or $P_{\mathrm{orb}}$.

The \simethod ~results for stellar \sdr ~are usually measured 
by using the angular
velocity difference 
\begin{eqnarray}
\Delta \Omega = 
\Omega_{\mathrm{max}}-\Omega_{\mathrm{min}}=
{{{2 \pi} \over {P_{\mathrm{min}}}}} - {{{2 \pi} \over {P_{\mathrm{max}}}}}
\label{omega}
\end{eqnarray}

\noindent
\citep[e.g.][their Table 1]{Bar05}.
For solar \sdr, 
these values are
$\Omega_{\mathrm{max}}=\Omega_{\mathrm{eq}}=2 \pi/P_{\mathrm{eq}}$
and
$\Omega_{\mathrm{min}}=\Omega_{\mathrm{pole}}=2 \pi/P_{\mathrm{pole}}$,
where $P_{\mathrm{eq}}\approx25^{\mathrm{d}}$ and $P_{\mathrm{pole}}\approx35^{\mathrm{d}}$ are
the periods at the equator and the pole, i.e.
the Sun has $P_{\mathrm{eq}} < P_{\mathrm{pole}}$.

\subsection{\simethod ~general results }
 \label{SIgeneral}

\noindent
In this section,
we discuss the \problemone ~in the {\it general} context.
Let us assume that the \simethod ~would be used to measure
SDR from the observed difference
between the periods of non-stationary starspots
($P_{\mathrm{act}}$)
and stationary starspots ($P_{\mathrm{rot}}$). 
We compute the stellar 
surface differential coefficient 
(see Eqs. \ref{rotationsun}, \ref{k_one} and \ref{k_two})
 values from
\begin{eqnarray}
|k| = | {
        {P_{\mathrm{act}}-P_{\mathrm{rot}}}
        \over
        {(P_{\mathrm{act}}+P_{\mathrm{rot}})/2}
        } |
\label{si_k}
\end{eqnarray}
\noindent
for the thirteen CABSs where the active
longitude period $P_{\mathrm{act}}$ was detected in 
\citet[][their Tables 1 and 3]{Jet17}.
The angular velocity range is
\begin{eqnarray}
|\Delta \Omega| = | 
          {{2 \pi}\over{P_{\mathrm{act}}}}-{{2 \pi}\over{P_{\mathrm{rot}}}}
                  |.
\label{si_omega}
\end{eqnarray}
\noindent
The results are given in the order of increasing $|\Delta \Omega|$
in Table \ref{falsediff}.
The lower and upper limits are
$|\Delta \Omega|=0.0002 {\mathrm{~rad~d^{-1}}}$ (DM~UMa) 
and
$|\Delta \Omega|=0.042{\mathrm{~rad~d^{-1}}}$ (V711~Tau),
respectively.
Our lower limit, $|\Delta \Omega| \approx 0$, agrees with the results of the
earlier \simethod ~studies 
because weak \sdr ~has been observed in several stars.
However, our $|\Delta \Omega|$ upper limit is about three times smaller than
in \citet[][Table 1: $0.14{\mathrm{~rad~d^{-1}}}$]{Bar05}
and over ten times smaller than in
\citet[][Table 2: $0.47{\mathrm{~rad~d^{-1}}}$]{Bal16}.

\citet{Bar05} and \citet{Bal16} give several
$|\Delta \Omega|$ values for the {\it same} star.
Even if the \twopmodel ~were correct, 
the {\simethod}s are not so accurate
that they would always give exactly the same
$|\Delta \Omega|$, or equivalently the same $P_{\mathrm{act}}$ and $P_{\mathrm{orb}}$,
value for the starspots of the {\it same} star.
This must have increased the scatter (i.e. the range)  of these
earlier published upper $|\Delta \Omega|$ limits.
In addition to the previously mentioned inversion uncertainties,
the identification of the {\it same} structure in temporally separated
different images of the {\it same} star is not always unambiguous
(see Sect \ref{mapinc}: map-\Inc).
Finally, there are several additional reasons
why our $|\Delta \Omega|=0.042{\mathrm{~rad~d^{-1}}}$ 
upper limit 
is certainly an underestimate, 
as will be explained later in 
Sects. \ref{indirect}.
In this {\it general} context, 
the earlier published \simethod ~$|\Delta \Omega|$ estimates of \sdr ~do
not contradict the idea that the observed light curves of 
CABSs and CASSs could be
interference of the two constant period
($P_{\mathrm{act}}$ and $P_{\mathrm{orb}} \approx P_{\mathrm{rot}}$) 
``real'' light curves of the \JHLarg.

\begin{figure} 
\resizebox{8.5cm}{!}{\includegraphics{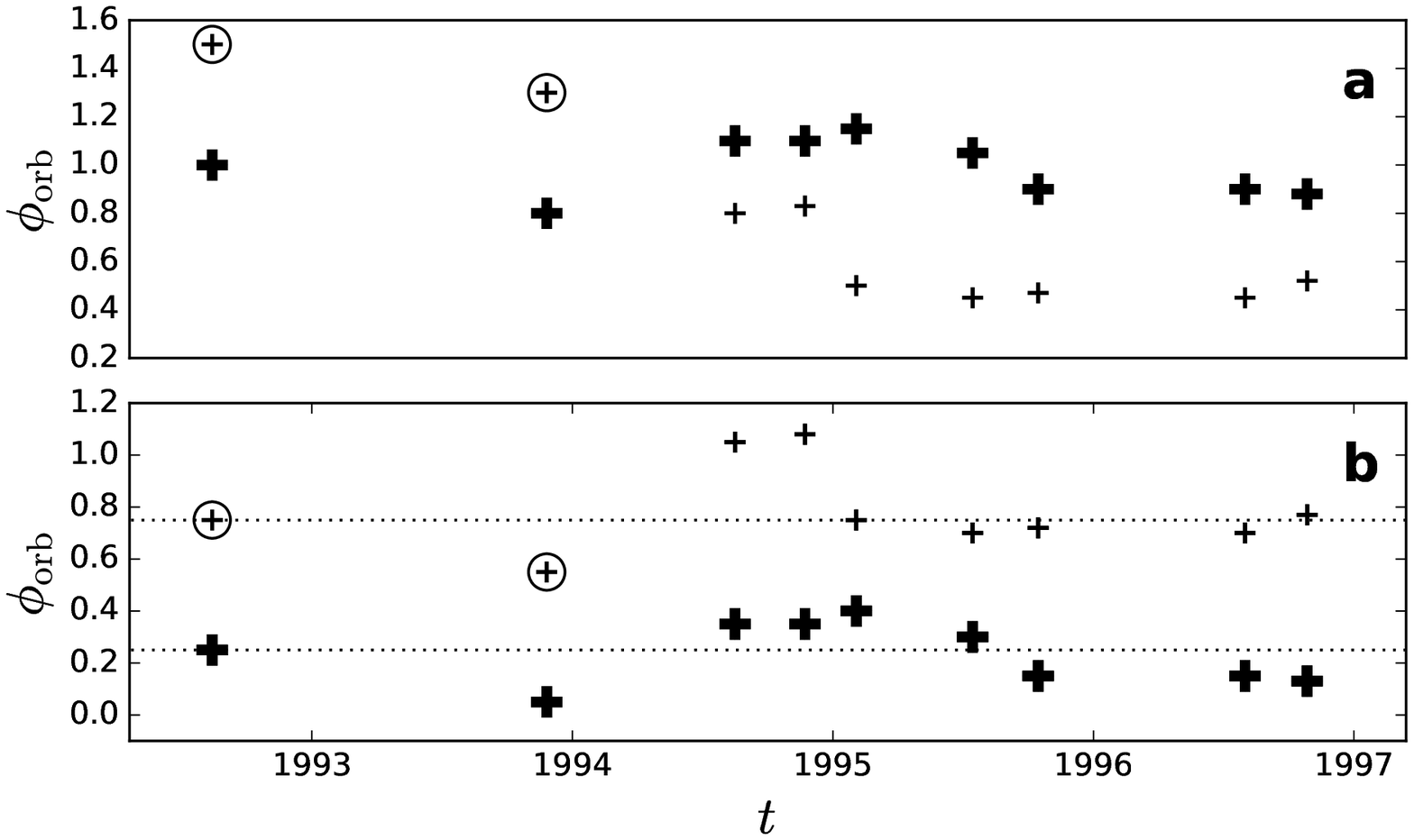}}
\caption{Orbital phases for starspots of II Peg detected with the \simethod. 
(a) Larger starspots (heavier crosses) and 
smaller starspots (lighter crosses) with the ephemeris of Eq. \ref{berephe}
\citep[][their Fig. 4]{Ber98B}.
The two highlighted lighter crosses are shifted one
orbital round downwards in the next panel.
(b) Same data with the ephemeris of Eq. \ref{jetephe}.
The horizontal dotted lines show the orbital phases 
$\phi_{\mathrm{orb}}=0.25$ and 0.75.
Otherwise as in (a).}
\label{false}
\end{figure}

\subsection{\simethod ~particular results }
\label{SIparticular}

Here, we discuss the \problemone ~in the
{\it particular} context of
the \simethod ~results presented
by \citet[][II Peg]{Ber98B} and \citet[][FK Com]{Kor00}.

\citet{Ber98B} published nine surface 
images of II Peg
between the years 1992 and 1997.
Their orbital phase ephemeris was
\begin{eqnarray}
{\mathrm{HJD~}}2449582.9268~+~6.724333{\mathrm{E}}.
\label{berephe}
\end{eqnarray}
\noindent
Their zero epoch was equal to that given
in \citet[][``Ac'' epoch in their Table 2]{Jet17}.
\citet{Ber98B} used the $P_{\mathrm{orb}}$  value 
of Eq. \ref{berephe} in the 
SI-method inversions.
Our Fig. \ref{false}a reproduces their phase plot 
for the larger and smaller starspots 
identified with the \simethod.
\citet{Ber98B} concluded that this diagram 
shows two active longitudes, 
and that a \flip ~event 
occurred in the year 1994.

\citet{Jet17} used the orbital period ephemeris
\begin{eqnarray}
{\mathrm{HJD~}}2449581.246~+~6.724333{\mathrm{E}}
\label{jetephe}
\end{eqnarray}
\noindent
for II Peg.
Their  $P_{\mathrm{orb}}$ was 
the same as in Eq. \ref{berephe}, 
but their zero epoch 
was $\Delta t = P_{\mathrm{orb}}/4 \equiv \Delta \phi_{\mathrm{orb}}=0.25$ 
earlier than in Eq. \ref{berephe} \citep[][``Ab'' epoch in their Table 2]{Jet17}.
One of their main results
was that all photometric minima of II Peg 
concentrated close
to $\phi_{\mathrm{orb}}=0.25$ between 
the years 1988 and 1997 
\citep[][their Fig. 27a]{Jet17}.
In other words, the stationary part of the light curve
($P_{\mathrm{orb}}=P_{\mathrm{rot}}$)
dominated during this nine year time interval
which overlaps the four year 
time interval of the \simethod ~images 
in \citet{Ber98B}.

Our Fig. \ref{false}b shows 
the phases of the same starspots as in Fig. \ref{false}a,
except that we use the ephemeris of Eq. \ref{jetephe}.
Unlike \citet{Ber98B}, 
we do not deliver evidence for a \flip ~event by
adding one extra orbital
round to the two first $\phi_{\mathrm{orb}}$ values of smaller starspots 
(Figs. \ref{false}ab: two highlighted lighter crosses).
The starspots of the first image in the year 1992 are exactly at 
$\phi_{\mathrm{orb}}=0.25$ and $\phi_{\mathrm{orb}}=0.75$,
respectively.
All larger starspots are at both sides of phase 
$\phi_{\mathrm{orb}}=0.25$.
Most of the smaller starspots concentrate close 
to $\phi_{\mathrm{orb}}=0.75$.
The simultaneous MLC of II Peg also confirms
that the main activity concentrated on the longitude
coinciding with the line connecting the
centres of the two binary components \citep[][Fig. 14d-e:
$\phi_{\mathrm{orb}}=0.25$]{Jet17}.
\citet{Ber98B} noted that the mean 
latitudes of these starspots discovered with
the \simethod ~remained 
approximately the same in all images.
This is a convincing {\it particular} case, where the 
\simethod ~detects a stationary 
starspot of II Peg
rotating with a constant period 
$P_{\mathrm{orb}}=P_{\mathrm{rot}}$
during a time interval of over four  years.

\citet{Kor00} published six surface images of FK Com between
the years 1994 and 1997. 
They used the period $P_{\mathrm{phot}}=2.^{\mathrm{d}}4002466\pm0.^{\mathrm{d}}0000056$
in the SI-method image inversions.
Comparison of their six images revealed the presence 
of long-lived starspots rotating with
a constant period $P_{\mathrm{SI}}=2.^{\mathrm{d}}4037 \pm 0.^{\mathrm{d}}0005$.
The dates of these images overlap the time interval of our photometry
between the years 1995 and 1997, except for the year 1994.
The thirteen first statistically independent
$t_{\mathrm{min,1}}$ epochs 
in our Table \ref{electric} 
(IND=1) are between the years 1995.43 and 1997.41.
The \kmethod ~analysis of these thirteen epochs 
gives 
$P_{\mathrm{act}}=2.^{\mathrm{d}}4035 \pm 0.^{\mathrm{d}}0005$.
The $P_{\mathrm{SI}}-P_{\mathrm{act}}$ difference
is only $0.^{\mathrm{d}}0002 \pm 0.^{\mathrm{d}}0007$.
The latitude range of the starspots detected with the SI--method
was between $45\degr$ and $70\degr$, 
and the migration of these starspots followed 
``solid body rotation''\citep{Kor00}.
In this {\it particular} case, the SI-method undoubtedly detects
a long-lived non-stationary starspot of FK Com rotating with a
constant period $P_{\mathrm{act}}$.

The starspots detected in
the above two {\it particular} \simethod ~studies 
by \citet{Ber98B} and \citet{Kor00} were long-lived and
rotated with a constant angular velocity,
as predicted by the \JHLarg.
This type of stability is quite common in surface images.
For example, the sequence 37 consecutive surface images
during 57 nights show that the strongest starspot in V711~Tau
appears to be nearly stationary in the rotating reference frame
\citep[][their Fig. 8: starspot ``A'']{Str00A}.
Our answer to the \problemone ~is
that the {\it general} and the {\it particular}
evidence strongly indicates that 
the $P_{\mathrm{act}}$ and $P_{\mathrm{rot}}\approx P_{\mathrm{orb}}$
periods of the \JHLarg ~have already been detected with
the \simethod.

\section{\lcmethod}
\label{lcmethodsect}

We give compact answers to the \problemtwo ~and the \problemthree ~in 
\yhteys{\paperii ~(Sect. 12)}.
These answers are based on the results presented in 
Sects. \ref{simulations}-\ref{periodinc}.

\subsection{\lcmethod ~parameters}
\label{LCparameters}

In the \lcmethod ~approach,
the most widely used period finding method 
is the Lomb-Scargle periodogram 
\citep[hereafter the \lsmethod,][]{Lom76,Sca82}.
It is equivalent to finding the best least squares
sinusoidal model for the data (hereafter \lsmodel).
The \lsmethod ~is a one-dimensional period
finding method because it determines only 
one period value for the data.
There are numerous other one-dimensional 
period finding methods 
that can search for more complex models in the data
\citep[e.g.][the \cpsmethod]{Leh11}.

The {\lcmethod s} usually
use the following approximation
for the law of solar \sdr
\begin{eqnarray}
P(b)={
      {P_{\mathrm{eq}}} 
       \over
      {1- k \sin^2 b}
      },
\label{rotationsun}
\end{eqnarray}
\noindent
where $b$ is the latitude,
$P_{\mathrm{eq}}$ is the rotation period at the equator
and $k$ is the \sdr ~coefficient
\citep[e.g.][solar $k=k_{\odot}=0.186$]{Hal90,Hal91A}.
A useful parameter is 
\begin{eqnarray}
h= \sin^2 (b_{\mathrm{max}}) -  \sin^2 (b_{\mathrm{min}}),
\label{latrange}
\end{eqnarray}
\noindent
where $b_{\mathrm{min}}$ and $b_{\mathrm{max}}$ are the
mimimum and maximum latitudes of starspot formation.
The largest possible latitude range,
$b_{\mathrm{min}}= 0 \degr$ and $b_{\mathrm{max}}= \pm 90\degr$, gives $h=1$.
All other latitude ranges give $h<1$.
This leads to the relation 
\begin{eqnarray}
k= {\Delta P \over {h P_{\mathrm{eq}}}},
\label{k_one}
\end{eqnarray}
\noindent
where $\Delta P= |P(b_{\mathrm{max}}) -  P(b_{\mathrm{min}})|=
P_{\mathrm{phot,max}}-P_{\mathrm{phot,min}}$ is the difference between the 
largest and smallest
observed photometric rotation period $P_{\mathrm{phot}}$
in any particular star.
Since $h\le1$, the {\lcmethod}s use the lower limit
\begin{eqnarray}
|k| \ge {{P_{\mathrm{phot,max}}- P_{\mathrm{phot,min}}} \over {P_{\mathrm{phot,mean}}}}
\label{k_two}
\end{eqnarray}
\noindent
for the absolute value of stellar \sdr ~coefficient,
where $P_{\mathrm{phot,mean}}$ 
is the mean of all observed values 
of $P_{\mathrm{phot}}$.
There are four uncertainties in this relation.
Firstly, 
$P_{\mathrm{phot,mean}}\approx (P_{\mathrm{phot,max}} +P_{\mathrm{phot,min}})/2$ 
must to be used 
as an approximate estimate for $P_{\mathrm{eq}}$ of Eq. \ref{k_one},
because it is not known whether $P_{\mathrm{phot,min}}$ or $P_{\mathrm{phot,max}}$
represents $P_{\mathrm{eq}}$.
Some {\lcmethod}s use $P_{\mathrm{eq}}=P_{\mathrm{phot,max}}$ \citep[e.g.][]{Rei13}.
Secondly, the sign of $k$ is unknown,
where $k >0$ represents solar and  $k < 0$ represents anti-solar \sdr.
Thirdly, the latitude range of starspot formation,
the value of $h$ in Eqs. \ref{latrange} and \ref{k_one}, is unknown.
Fourthly, it is not known if the full range of period
changes has already been observed,
or does the observed 
$P_{\mathrm{phot,max}}-P_{\mathrm{phot,min}}$ difference 
represent an underestimate of the real range.

Another approach for measuring the range of changes for
$n$ photometric rotation periods $P_{\mathrm{phot,i}}$ 
\citep[e.g.][]{Leh16} is to use the weighted mean
\begin{eqnarray}
P_{\mathrm{w}}= {
             {\sum_{i=1}^n w_i P_{\mathrm{phot,i}}}
             \over
             {\sum_{i=1}^n w_i}
             }
\label{Pw}
\end{eqnarray}
and the weighted standard deviation
\begin{eqnarray}
\Delta P_{\mathrm{w}}= 
\sqrt{
             {
             {\sum_{i=1}^n w_i (P_{\mathrm{phot,i}}-P_{\mathrm{w}})^2}
             \over
             {\sum_{i=1}^n w_i}
             }
      },
\label{ePw}
\end{eqnarray}

\noindent
where 
$\sigma_{P_{\mathrm{phot,i}}}$ is the error of $P_{\mathrm{phot,i}}$
and the weights are $w_i=\sigma_{P_{\mathrm{phot,i}}}^{-2}$.  
Equal weights $w_i=1$ give $P_{\mathrm{w}}=m_{\mathrm{P}}$ 
and
$\Delta P_{\mathrm{w}}=\sigma_{\mathrm{P}}$,
where $m_{\mathrm{P}}$ and $\sigma_{\mathrm{P}}$ are 
the mean and standard deviation
of all observed $P_{\mathrm{phot,i}}$ values.
In this case,
the ``three sigma'' upper limit for the
$P_{\mathrm{phot,i}}$ changes is 
\begin{eqnarray}
Z={{6 \Delta P_{\mathrm{w}}}\over{P_{\mathrm{w}}}} 
= { {6 \sigma_{\mathrm{P}}}\over{m_{\mathrm{P}}}}.
\label{zvalue}
\end{eqnarray}
\noindent
The relation
\begin{eqnarray}
|k| \approx Z/h 
\label{k_three}
\end{eqnarray}
\noindent
is valid, if 
$P_{\mathrm{phot,max}}-P_{\mathrm{phot,min}} \approx 6 \Delta P_{\mathrm{w}}$
\citep[][]{Jet00}.

\subsection{\lcmethod ~results}
 \label{lcmethodresults} 

The idea that the value of $P_{\mathrm{phot}}$ is connected
to the stellar rotation period $P_{\mathrm{rot}}$ is logical.
In the \sdr ~context, 
the observed value of $P_{\mathrm{phot}}$ could tell
something about the latitude of the spot.
If the {\it observed} $P_{\mathrm{phot}}$ has changed,
it is logical to assume that the latitude of the spot has changed.
Thus, the range of  $P_{\mathrm{phot}}$ changes
could measure \sdr.

\citet{Hal91A} applied the \lcmethod ~to 
the photometry of 277 late-type stars. 
He concluded that the stellar \sdr ~correlates 
strongly with $P_{\mathrm{phot}}\approx P_{\mathrm{rot}}$.
It decreases when $P_{\mathrm{phot}}$ decreases.
He showed that
\sdr ~was so weak 
in some rapidly rotating stars that
it approached solid-body rotation.
This general relation between $P_{\mathrm{phot}}$ and
\sdr ~has been amply confirmed by subsequent
\lcmethod ~studies of much larger 
samples \citep[e.g.][24~124 stars]{Rei13}.
We have also formulated our own {\lcmethod}s, 
and used them to measure stellar \sdr ~
\citep[e.g.][the TSPA-method and the \cpsmethod]{Jet99,Jet99A,Leh11,Leh16}.
Considering all these findings, 
our \JHLarg ~runs into its next problem:
what is the answer to the \problemtwo.
Note that we have very good reasons for emphasizing the words
{\it many}, {\it different},
{\it observed} and {\it same} in this particular question, 
as will become evident in the
next Sects. \ref{simulations}-\ref{periodinc}.

\begin{figure}
\resizebox{8.5cm}{!}{\includegraphics{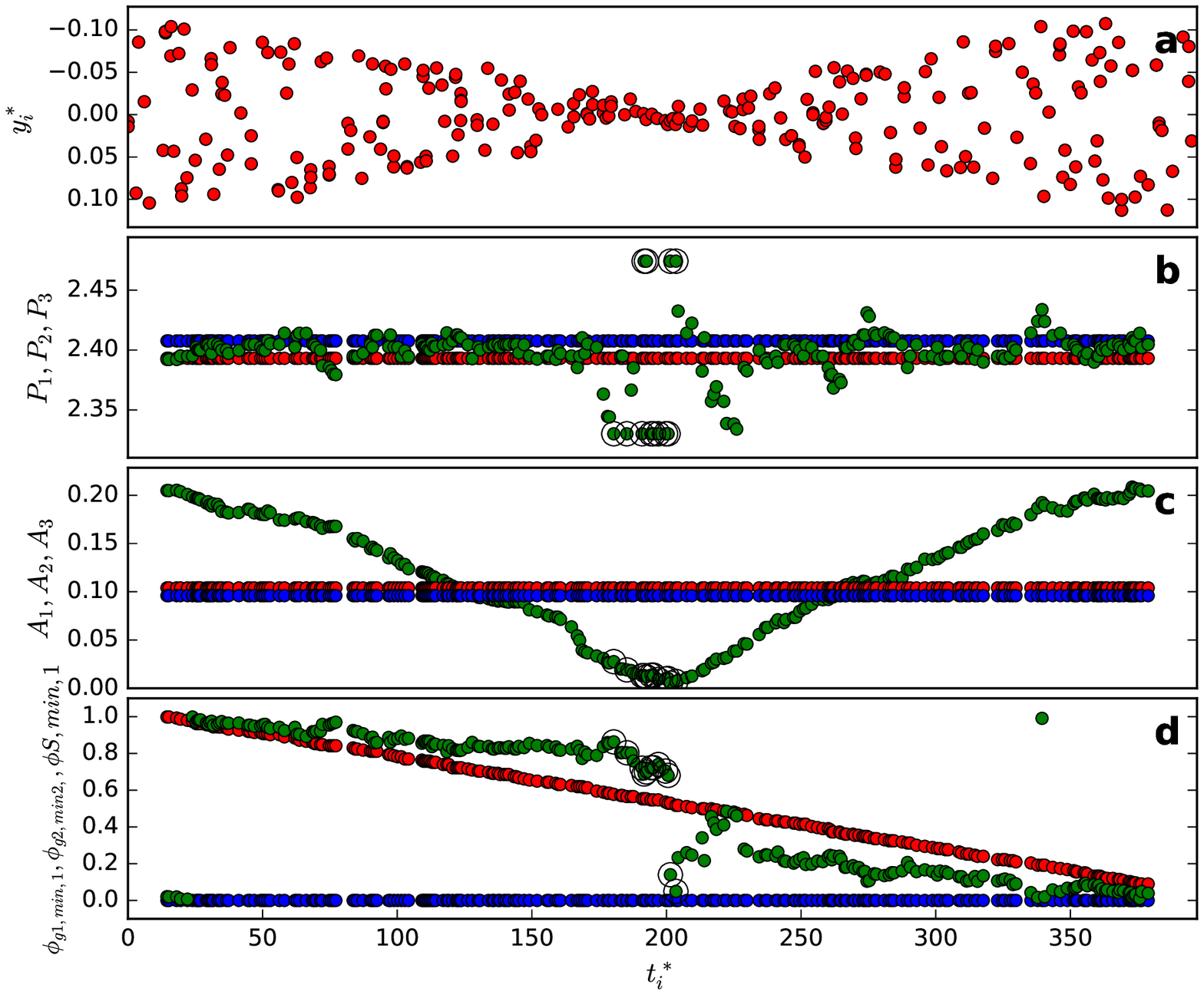}}
\caption{
(a) Simulated $y^*_i$ data of Eq. \ref{simulateddata} for
equal amplitudes $a_1=a_2=0.^{\mathrm{m}}05$ in Eqs. \ref{sone} and \ref{stwo}.
(b) Periods $P_1=2.^{\mathrm{d}}39313$ and $P_2=2.^{\mathrm{d}}40762$ of the 
\twopmodel ~are denoted with red and blue circles. 
The green circles show the periods $P_3$ detected with the \lsmethod.
The large black circles around some green circles highlight the cases,
where the highest $z_{\mathrm{LS}}$
periodogram value is in the beginning or 
in the end of the tested period interval.
(c) Amplitudes $A_1=2a_1$ and $A_2=2a_2$ of the \twopmodel ~are denoted
with red and blue circles. 
Note that these red and blue symbols are slightly shifted apart,
because they would otherwise overlap.
The green circles show the peak to peak amplitudes $A_3$ 
of the \lsmodel ~sinusoids. 
The same models as in "b" are highlighted again.
(d) The red and blue circles show 
the light curve minimum phases of $g_1(t)=s_1(t)$ and $g_2(t)=s_2(t)$ 
with $P_2$ in Eq. \ref{phaseone}.
The green circles show the light curve minima of the \lsmethod ~
sinusoids.  The highlighted models are the same as in "bc".
}
\label{figsimulated1}
\end{figure}

\begin{figure}
\resizebox{8.5cm}{!}{\includegraphics{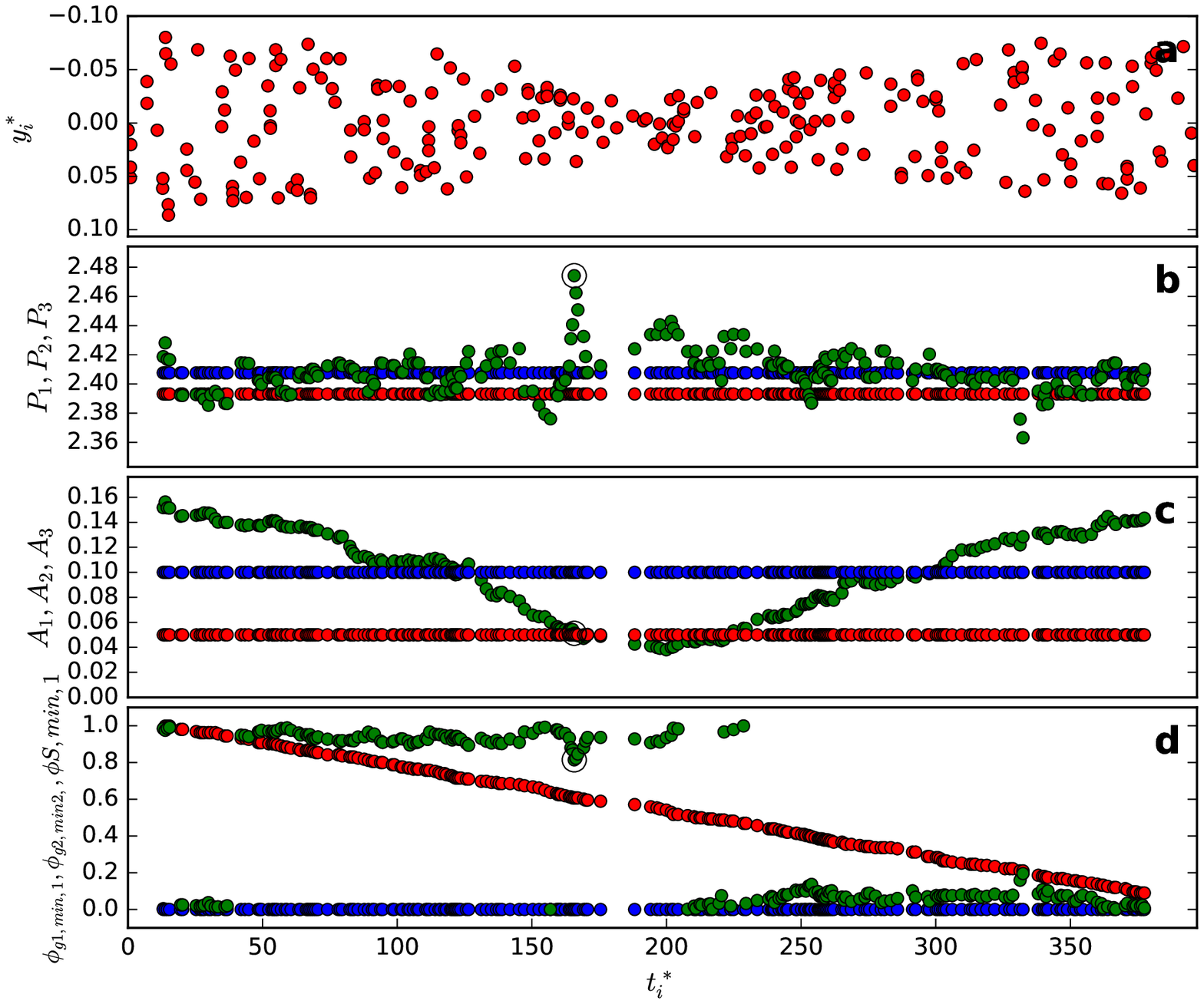}}
\caption{Simulated $y^*_i$ data of Eq. \ref{simulateddata} for
unequal amplitudes $a_1=0.^{\mathrm{m}}025$ in Eq. \ref{sone} 
and $a_2=0.^{\mathrm{m}}050$ in \ref{stwo}, 
otherwise as in Fig. \ref{figsimulated1}.} 
\label{figsimulated2}
\end{figure}

\subsection{Simulated \twopmodel ~data }
\label{simulations} 

Here, we assume that the \twopmodel ~is correct
and we simulate artificial photomeric data of two different cases.
The model used in creating the simulated data 
is the sum of two sinusoidal light curves 
\begin{eqnarray}
s_1(t) & = & a_1\sin{(2 \pi f_1 t)} \label{sone}\\
s_2(t) & = & a_2\sin{(2 \pi f_2 t)}. \label{stwo}
\end{eqnarray}
This model has
$s_1(t)=g_1(t)$ in Eq. \ref{gone},
$s_2(t)=g_2(t)$ in Eq. \ref{gtwo}
and
$s(t)=s_1(t)+s_2(t)=g_{\mathrm{C}}(t)$ in Eq. \ref{gtwomodel}.
The peak to peak amplitudes are $A_1=2a_1$ and $A_2=2a_2$.
The frequencies are fixed to $f_1=P_1^{-1}$ and $f_2=P_2^{-1}$,
where the periods are 
$P_1=2.^{\mathrm{d}}39313$ and
$P_2=2.^{\mathrm{d}}40762$. 
These period values are taken from the ephemerides
of \yhteys{Eqs. 20 and 21 of \paperii}.
We compute the lap cycle period   
\begin{eqnarray}
P_{\mathrm{lap}}=|P_1^{-1}-P_2^{-1}|^{-1}=|f_1-f_2|^{-1}\approx 398^{\mathrm{d}}
\label{lapcycle}
\end{eqnarray}
of these periodicities \citep[][their Eq. 4]{Jet17}.
The interference pattern of $s(t)=s_1(t)+s_2(t)$ 
is repeated 
regularly after every $\Delta T = P_{\mathrm{lap}}$.
Hence, the time span of the simulated data segment
is fixed to 
$\Delta T = P_{\mathrm{lap}}$.
We use multiples of sidereal day 
$P_{\mathrm{sid}}=0.^{\mathrm{d}}99726958$ 
to create $n^*=250$ time points
\begin{eqnarray}
t^*_i= i^* P_{\mathrm{sid}}+\delta t^*_i,
\nonumber
\end{eqnarray}
where $i^*$ are a random sample of integers
$1 \le i^* \le 398$. Any particular 
integer $i^*$ value is used as many times
as the random selection favours it.
The simulated random shifts $\delta t^*_i$ at both sides of
the mid-nights $i^* P_{\mathrm{sid}}$ are evenly
distributed between $-0.^{\mathrm{d}}2$ and $0.^{\mathrm{d}}2$.
The simulated data are
\begin{eqnarray}
y^*_i=s(t^*_i)+\epsilon^*_i,
\label{simulateddata}
\end{eqnarray}
\noindent
where $\epsilon^*_i$ are a random sample 
of $n^*=250$ residuals drawn from a Gaussian distribution with
a zero mean and a standard deviation 
of $0.^{\mathrm{m}}008$.

Inside the simulated segment,
we select datasets containing all $y^*_{\mathrm{i}}$
that are within a sliding window of 
$\Delta T=30^{\mathrm{d}}$.
We apply the \lsmethod ~to all those datasets that contain at 
least $n=12$ values of $y^*_i$.
Similar sliding window one-dimensional period
analysis has been applied to real photometry,
e.g. by \citet[][the \lsmethod]{Dis16} or \citet[][the \cpsmethod]{Leh16}.
The maximum and minimum values of tested $f_3=P_3^{-1}$ frequencies
of the \lsmodel ~are
\begin{eqnarray}
f_{\mathrm{min}} =  (1-a)f_{\mathrm{mid}}, & &  
f_{\mathrm{max}} =  (1+a)f_{\mathrm{mid}},
\label{fsimraja}
\end{eqnarray}
where
$P_{\mathrm{mid}}=(P_1+P_2)/2$,
$f_{\mathrm{mid}}=P_{\mathrm{mid}}^{-1}$,
and $a=0.03$.
This is the $\pm 3$\% range 
at both sides of $f_{\mathrm{mid}}$.

\subsubsection{$a_1=a_2$ simulation (\testone)
 \label{test1} }

In the first case, we use equal amplitudes  
$a_1=a_2=0.^{\mathrm{m}}05$ in Eqs. \ref{sone} and \ref{stwo}. 
We will hereafter refer to these simulations as \testone.
The results are shown in Fig. \ref{figsimulated1}.

The amplitude of the simulated data decreases to zero during
the first half of the segment and then increases back to its original
level during the last half
(Fig. \ref{figsimulated1}a).
This pattern is repeated during every lap cycle 
$\Delta T = P_{\mathrm{lap}}$ (Eq. \ref{lapcycle}),
because the interference of these two sinusoids is regular.
The periods $P_1$ and $P_2$ of the simulated model
do not change (Fig. \ref{figsimulated1}b: red and blue circles).
However, the periods detected with the \lsmethod ~show
large variation  (Fig. \ref{figsimulated1}b: green circles),
especially when the amplitude of the simulated data approaches zero
in the middle of the segment.
There are cases where
the highest \lsmethod ~periodogram $z_{\mathrm{LS}}$ value is at the
end of the tested period interval
(Fig. \ref{figsimulated1}b: highlighted green circles).
This is a clear sign of that the tested period interval
is too narrow for this particular period finding method.
The amplitudes $A_1$ and $A_2$ of the \twopmodel ~remain
unchanged  (Fig. \ref{figsimulated1}c: red and blue circles).
The peak to peak amplitudes $A_3$ of the sinusoidal {\lsmodel}s
vary between $0.^{\mathrm{m}}0$ and $0.^{\mathrm{m}}2$
(Fig. \ref{figsimulated1}c: green circles).
The phases of the light curve minima of $s_2(t)$ are stable
with $P_2$ in Eq. \ref{phaseone} (Fig. \ref{figsimulated1}d: blue circles),
while those of $s_1(t)$ show regular linear migration 
(Fig. \ref{figsimulated1}d: red circles).
Finally, the minima of the sinusoidal models detected
with the \lsmethod ~show minor linear changes and
an abrupt 0.5 phase shift in the middle
of the segment (Fig. \ref{figsimulated1}d: green circles).
This phase shift occurs when the amplitude $A_3$ approaches zero,
and this event coincides with the cases when no periodicity is detected with the 
\lsmethod ~(Figs. \ref{figsimulated1}bcd: highlighted green circles).
We emphasize that this half a rotation phase shift occurs in
{\it every} simulated artificial data sample of Eq. \ref{simulateddata}
when $a_1=a_2$.

\subsubsection{$a_1<a_2$ simulation (\testtwo) }
\label{test2} 

In the second case, the amplitudes are unequal:
$a_1=0.^{\mathrm{m}}025$ in Eq. \ref{sone}
and 
$a_2=0.^{\mathrm{m}}05$ in Eq. \ref{stwo}. 
These simulations are hereafter referred to as \testtwo.

The amplitude of the simulated data first decreases 
from $0.^{\mathrm{m}}15$ to $0.^{\mathrm{m}}05$,
and then increases back to  $0.^{\mathrm{m}}15$
(Fig. \ref{figsimulated2}a).
The periods $P_1$ and $P_2$ do not, of course, change
(Fig. \ref{figsimulated2}b: red and blue circles).
There is large variation in the $P_3$ periods detected with the \lsmethod 
~(Fig. \ref{figsimulated2}b: green circles).
This variation is largest in the middle of the segment
when the amplitude is lowest.
The amplitudes $A_1$ and $A_2$ of 
the \twopmodel ~do not change (Fig. \ref{figsimulated2}c: red and blue circles).
The peak to peak amplitudes $A_3$ of the {\lsmodel}s vary
regularly between $0.^{\mathrm{m}}05$ and $0.^{\mathrm{m}}15$
(Fig. \ref{figsimulated2}c: green circles).
The same phases
of the light curve minima of $s_1(t)$ and $s_2(t)$
as in the previous Fig. \ref{figsimulated1}d 
are shown again in Fig. \ref{figsimulated2}d (red and blue circles).
The phases of the minima of the sinusoidal {\lsmodel}s fluctuate
(Fig. \ref{figsimulated2}d: green circles).
The cases when the \lsmethod ~detects no periodicity
do not occur as often as in \testone,
because the amplitude does not decrease 
to zero in the middle of the segment 
(Figs. \ref{figsimulated2}bcd: one highlighted circle).

\begin{figure}
\resizebox{8.5cm}{!}{\includegraphics{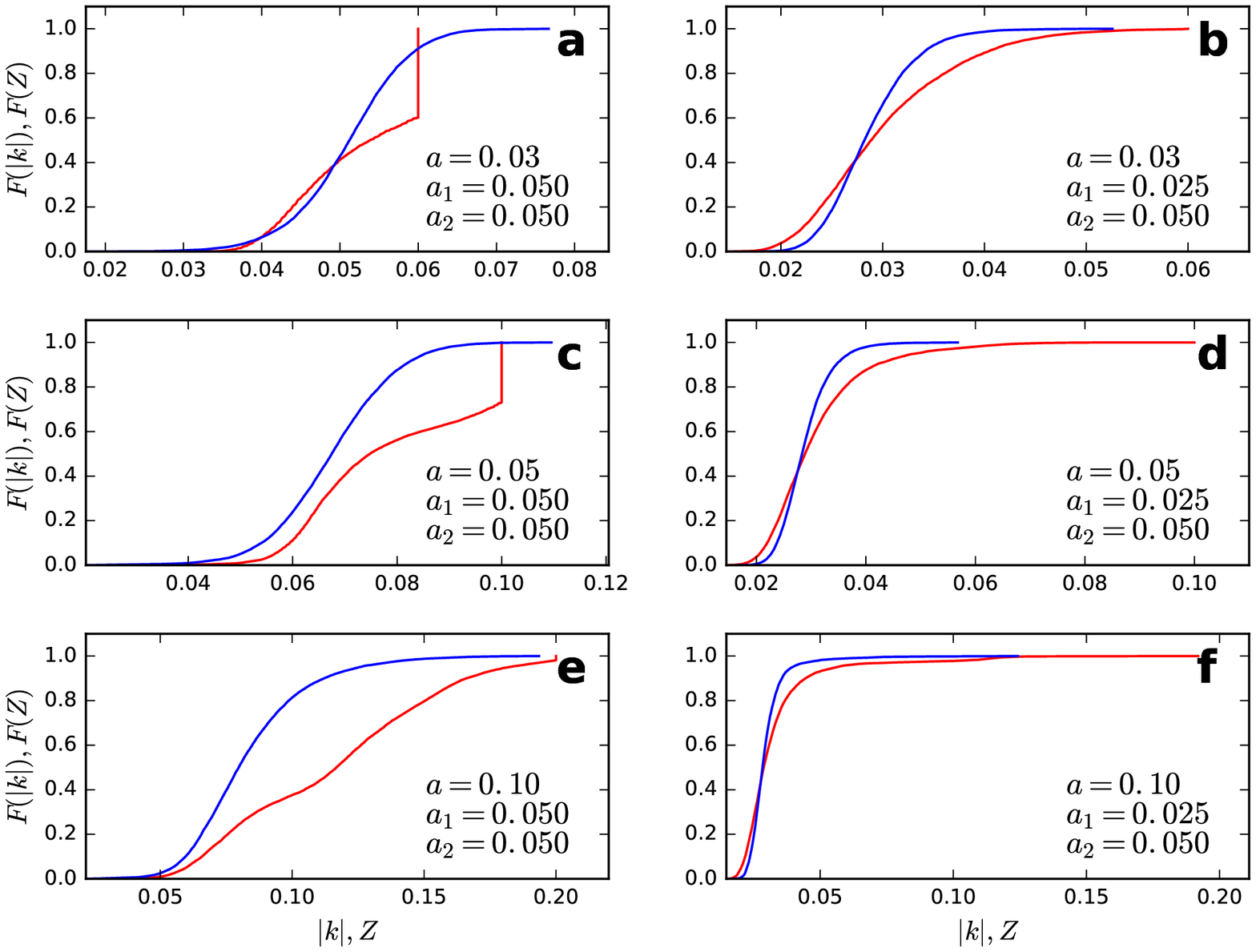}}
\caption{(a) Cumulative distribution functions 
$F(|k|)$ (red) and $F(Z)$ (blue) in 10~000 simulations of 
\testone ~with $a=0.03$ in Eq. \ref{fsimraja}
(b) same as in ``a'' for \testtwo.
(c) $a=0.05$ in Eq. \ref{fsimraja}, otherwise as in ``a''.
(d) $a=0.05$ in Eq. \ref{fsimraja}, otherwise as in ``b''.
(e) $a=0.10$ in Eq. \ref{fsimraja}, otherwise as in ``a''.
(f) $a=0.10$ in Eq. \ref{fsimraja}, otherwise as in ``b''.}
\label{hammer}
\end{figure}

\begin{figure}
\resizebox{8.5cm}{!}{\includegraphics{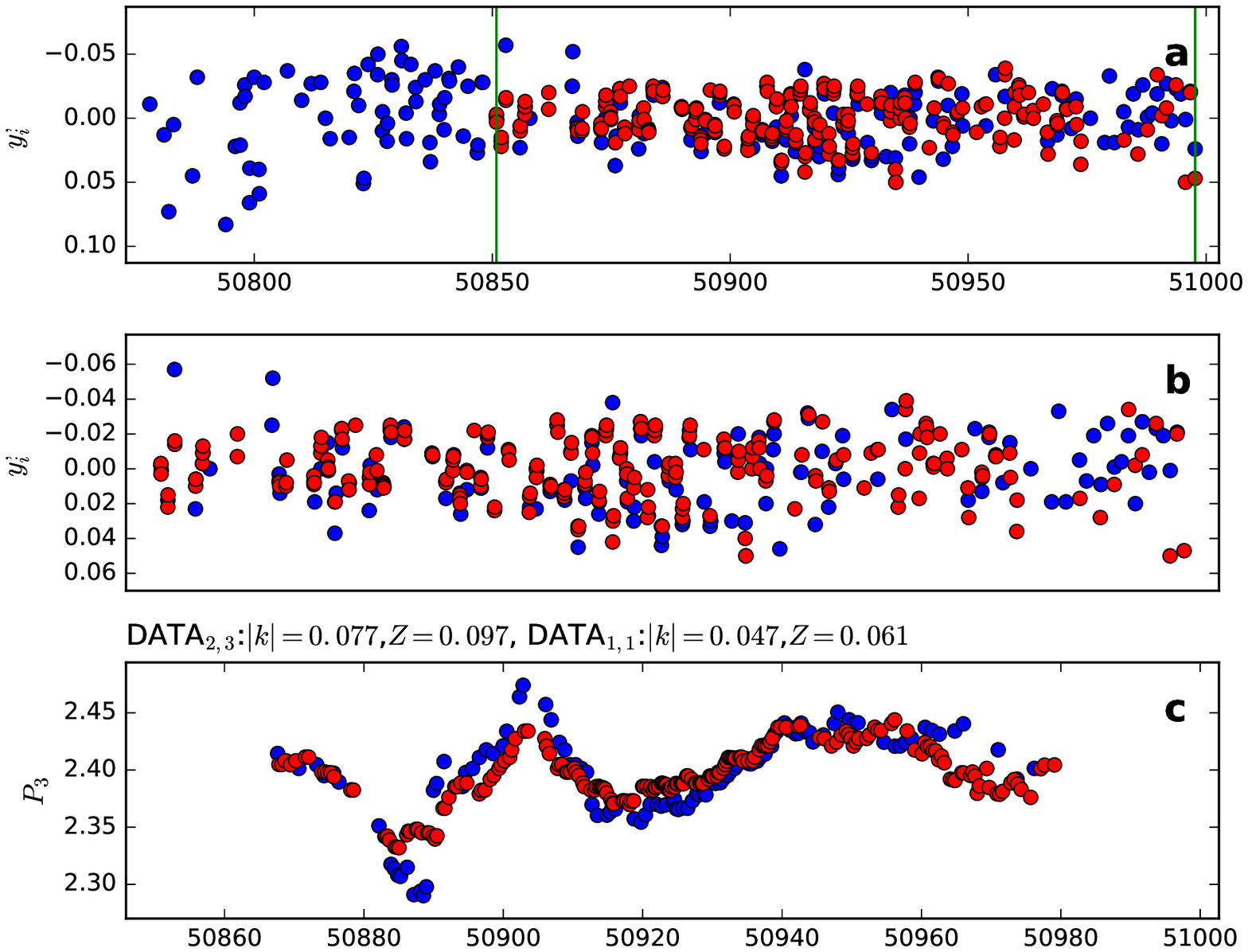}}
\caption{(a) Overlapping \Data{2}{3} and \Data{1}{1} 
photometry (red and blue filled circles).
Vertical green lines show the beginning and the end
of fully overlapping observations.
(b) Fully overlapping photometry.
(c) Red and blue filled circles
denote $P_3$ periods detected 
from the fully overlapping
photometry with 
the \lsmethod.
The $|k|$ and $Z$ results are 
given above this panel. }
\label{incompatibleP2}
\end{figure}

\subsection{Incompatibility} 
\label{incomp} 

The \testone ~and \testtwo ~simulations
reveal something that we will hereafter 
refer to as ``\Inc''.
If the \JHLarg ~is true,
this \Inc ~manifests itself in the {\it observed}
light curve periods, amplitudes and minima 
determined with any one-dimensional period
finding method, like
e.g. the \lsmethod ~or the \cpsmethod.

The amplitudes $A_1$ and $A_2$ of the simulated
light curves are constant in the above
mentioned \testone ~and 
the \testtwo ~simulations.
All \Inc ~effects described in the next 
Sects. \ref{periodinc}-\ref{mapinc}
are even stronger in the real data,
because these $A_1$ and $A_2$ amplitudes are changing.
These effects are nicely illustrated in the
periodogram of all data which resembles
a patchwork quilt
\yhteys{(see \paperii: Fig. A70 discussed in Sect. 9)}.

\subsubsection{Period-\Inc} 
\label{periodinc} 

The \testone ~and \testtwo ~simulations
both show that the \lsmethod ~detects 
spurious $P_3$ periods and spurious 
$P_3$ changes although 
the \twopmodel ~periods $P_1$ 
and $P_2$ do not change. 

The $P_3$ period results in Fig. \ref{figsimulated1}b
represent only one particular simulated 
\testone ~case 
where $a=0.03$ in Eq. \ref{fsimraja},
and $a_1=a_2=0.050$ 
in Eqs. \ref{sone} and \ref{stwo}.
We simulated 10~000 such cases having this
same $a$, $a_1$ and $a_2$ combination.
Each case gave us one estimate of
$|k|$ (Eq. \ref{k_two}) 
and $Z$ (Eq. \ref{zvalue}).
The cumulative distribution
functions $F(|k|)$ and $F(Z)$ 
of these parameters
are shown in Fig. \ref{hammer}a
(red and blue continuous lines).
The parameter ranges are approximately 
$0.035\le |k| \le 0.06$ and $0.03 \le Z \le 0.07$.
A steep rise of $F(|k|)$ occurs
at $2a=0.06$.
The probability for $F(|k| \ge 2a|)$
is nearly 0.5.
The $|k|$ values can not 
exceed this $2a$ limit,
because the maxima and
the minima of detected $P_3$ values have to be
within the tested period interval 
(Eq. \ref{fsimraja}).
In nearly half of the 10~000 cases,
the \lsmethod ~detects $P_3$ periods
over the whole tested period interval
$(1 \pm a) P_{\mathrm{mid}}$.
The shape of $F(Z)$ resembles that of
a Gaussian cumulative distribution function.

The results for 10~000 simulated 
\testtwo ~cases with
$a=0.03$, $a_1=0.025$ and $a_2=0.050$ 
are shown in Fig. \ref{hammer}b.
The approximate ranges are 
$0.02\le |k| \le 0.06$ and 
$0.02 \le Z \le 0.05$.
The minimum peak to peak amplitude 
of the simulated $s(t)$
light curve is $A_3=2 a_1=0.050$
in these \testtwo ~simulations.
Since $A_3$ does not decrease to zero,
as in \testone, 
the \lsmethod ~succeeds in detecting
periodicities that are
within the  tested period interval.

The combination 
$a=0.05$, $a_1=0.050$ and $a_2=0.050$ 
(Fig. \ref{hammer}c) shows what happens in
\testone ~with a longer tested $\pm 5$\% 
period interval.
The $|k|$ and $Z$ ranges increase,
and the ``expected'' steep $F(|k|)$ rise 
occurs at $2a=0.10$.
The \testtwo ~combination 
$a=0.05$, $a_1=0.025$ and $a_2=0.050$ 
shows no steep $F(|k|)$ rise 
at $|k|=2a$ (Fig. \ref{hammer}d).
The results for $a=0.10 \equiv \pm 10$\% 
are shown in Figs. \ref{hammer}ef.
The \lsmethod ~keeps on finding $P_3$
periods over the whole period interval
in \testone ~simulations,
because the $|k|$ maximum is at $2a=0.20$.
The general result is that the
$|k|$ and $Z$ ranges increase when 
the tested range, $a$ in Eq. \ref{fsimraja}, increases.

The simulations shown in Figs. \ref{hammer}a--f 
also reveal that the
relation $|k| \approx Z/h$ of Eq. \ref{k_three}
is a poor approximation.
This is no surprise,
because $|k|$ is computed from
{\it two} individual $P_3$ values, 
while $Z$ is computed 
from {\it all} $P_3$ values.

If the \JHLarg ~is correct,
our simulations also predict 
that two observers {\it observing} 
the light curve of {\it same} star 
during the {\it same} time interval
will get {\it different}
$P_3$, $|k|$ and $Z$ values with the \lsmethod.
We show the temporally overlapping 
\Data{2}{3} and \Data{1}{1}
photometry in Fig. \ref{incompatibleP2}a. 
The fully overlapping data are shown in  
Fig. \ref{incompatibleP2}b.
We divide these fully 
overlapping data into 
subsets using a sliding
window of 30 days,
and analyse
all subsets having at least 12 observations
with the \lsmethod 
~(Eq. \ref{fsimraja}: $a=0.05$).
In Fig. \ref{incompatibleP2}c,
we show the $P_3$, $|k|$ and $Z$ results for the fully 
overlapping data of segments \Data{2}{3} and \Data{1}{1}. 
The respective results for all temporally overlapping 
pairs of segments
are given in Table \ref{compsegments}.
These results confirm the above prediction.

If the \JHLarg ~is true,
no-one ever gets the same \lsmethod ~analysis results
from the simultaneous 
photometry of the same star.
This general result  
applies to all one-dimensional
period finding methods.
All the above findings are amazing 
considering that 
the underlying \twopmodel, 
$g_{\mathrm{C}}(t)=s(t)=s_1(t)+s_2(t)$,
does not change at all.
Since this period \Inc ~solves the \problemtwo,
it is easy to the answer 
the \problemthree ~(\yhteys{see \paperii: Sect. 12}).

\begin{table}
\caption{$|k|$ and $Z$ for fully overlapping real observations.
Columns 1 and 4 give the temporally overlapping segments.
The other columns give the $|k|$ and $Z$ values
for fully overlapping data.}
\begin{center}
\begin{tabular}{lcclcc} 
\hline
           &  $|k|$  & $Z$   &              &   $|k|$ & $Z$   \\
\hline
\Data{2}{3}: & 0.077 & 0.097 & \Data{1}{1}:  & 0.047 & 0.061 \\
\Data{2}{4}: & 0.100 & 0.141 & \Data{1}{2}:  & 0.063 & 0.085 \\
\Data{2}{5}: & 0.022 & 0.032 & \Data{1}{3}:  & 0.012 & 0.016 \\
\Data{2}{6}: & 0.014 & 0.018 & \Data{1}{4}:  & 0.010 & 0.012 \\
\Data{2}{7}: & 0.079 & 0.068 & \Data{1}{5}:  & 0.070 & 0.096 \\
\Data{2}{8}: & 0.051 & 0.062 & \Data{1}{6}:  & 0.025 & 0.032 \\
\Data{2}{9}: & 0.088 & 0.091 & \Data{1}{7}:  & 0.048 & 0.038 \\
\hline
\end{tabular}
\end{center}
\label{compsegments}
\end{table}

\subsubsection{Amplitude-\Inc}
\label{amplitudeinc} 

The {\it observed} light curve amplitudes $A_3$ in Figs. 
\ref{figsimulated1}c and
\ref{figsimulated2}c do not certainly tell
anything about the {\it real} light curve
amplitudes $A_1$ and $A_2$ of the \twopmodel.
We will hereafter refer to this problem 
as ``amplitude-\Inc''.
It is common for all one-dimensional
period finding methods.

The period analysis of the observed amplitudes 
$A_3$ can sometimes be used to determine the numerical
value for the lap cycle period $P_{\mathrm{lap}}$ of
Eq. \ref{lapcycle}. 
These $P_{\mathrm{lap}}$ cycles were clearly present in the
observed $A_3$ amplitudes of thirteen CABSs
\citep[][their Figs. 15-27]{Jet17}.
The numerical $P_{\mathrm{lap}}$ value can also be used to
indirectly estimate other \twopmodel ~parameters
(see Sect. \ref{indirect}).

\subsubsection{Minima-\Inc}
\label{minimainc} 

The one-dimensional period finding methods also suffer
from something that we will hereafter refer
to as ``minima-\Inc''.
If we would analyse the light curve minima epochs of
Figs. \ref{figsimulated1}d and  \ref{figsimulated2}d
with the \kmethod,
we would not detect the $P_1$ period 
or the $P_2$ period.
We could
detect $P_2$ when $A_1=0$ (\secondcase)
or $P_1$ when $A_2=0$ (\thirdcase).
Unfortunately, the observed light curve does not give
us any unambiguous information of when $A_1=0$ or 
when $A_2=0.$
Thus, the next logical step is to begin to search for
the answers to the \problemfour, the \problemfive ~and the \problemsix.

\subsubsection{Map-\Inc} 
\label{mapinc} 

If the \JHLarg ~is true,
the non-stationary and stationary
starspots migrate with different  $\Omega$ angular velocities.
In this case, the comparison of
surface images from different
years can  lead to false identifications of
the non-stationary ($P_{\mathrm{act}}$)
and
the stationary ($P_{\mathrm{orb}}$ or $P_{\mathrm{rot}}$)
starpots.
In general,
there is no unique period in the $g_{\mathrm{C}}(t)$ model
for comparing the
longitudinal difference between two starspots, say A and B,
over a longer period of time.
It is also uncertain, 
if A moves forwards or backwards with respect to B.
The \twopmodel ~predicts that
one can reliably compare 
two surface images of the {\it same} star
with the {\it same} constant period
only if the time difference of these surface images
is much shorter than $P_{\mathrm{lap}}/2$.
For example, the lap cycle period of FK Com
is only $P_{\mathrm{lap}}=398^{\mathrm{d}}$.
This is comparable to the typical one year gap
between observing seasons.
We will hereafter refer to this problem as ``map-\Inc''.
This effect resembles the shuffling of cards.
The current order of cards (i.e. longitudes of starspots A and B) does
not give any information about the order before the shuffling. 
This effect has certainly been overlooked or underestimated
in earlier \simethod ~studies,
like 
\citet[][EI Eri: eleven years]{Was09},
\citet[][II Peg: six years]{Lin11}
or
\citet[][FK Com: 13 years]{Hac13}.
This map \Inc ~must have caused numerous 
misidentifications of the rotating
structures, because there are no unique phases or
longitudes in the constantly changing \twopmodel. 
Such misidentifications have increased
the range of $|\Delta \Omega|$ estimates,
e.g. in \citet{Bar05} or \citet{Bal16},
which were discussed earlier in Sect \ref{simethodsect}.

\subsection{Dead end}
\label{deadend}

\citet{Jet17} assumed that 
the \kmethod ~analysis of the 
$t_{\mathrm{min,1}}$ minima epochs of Table \ref{electric}
would give an unambiguous $P_{\mathrm{act}}$ value for FK Com
(the \problemfour).
The next step in their reasoning was to use
this $P_{\mathrm{act}}$ value in solving an unambiguous
$P_{\mathrm{rot}}$ value for FK Com.
However, we have just shown 
that {\it if} the $t_{\mathrm{min,1}}$ minima epochs are determined
with any one-dimensional period analysis method,
{\it then} the minima-\Inc ~prevents a unique $P_{\mathrm{act}}$ solution.
Any particular $t_{\mathrm{min,1}}$ value 
in our Table \ref{electric} may, 
or may not, be connected to
$P_{\mathrm{act}}$ and/or $P_{\mathrm{rot}}$
(see Sect. \ref{indirect}: 
\migrone,
\migrtwo ~and
\migrthree  ~alternatives).
This problem could be at least partly solved,
if we knew the time intervals {\it when} 
the interference caused by the other periodicity is absent,
and {\it only} the
$P_{\mathrm{act}}$ or $P_{\mathrm{rot}}$ period
totally dominates the light curves of FK Com
(\secondcase ~or \thirdcase).
However, 
we have no way of knowing when takes place.

We could try to use the known orbital 
period $P_{\mathrm{orb}} \approx P_{\mathrm{rot}}$ to solve $P_{\mathrm{act}}$,
if FK Com were a binary, 
like the 14 CABSs studied in \citet{Jet17}.
This {\it certainly known} $P_{\mathrm{rot}}$ could then be used
to solve $P_{\mathrm{act}}$ in a similar way,
as $P_{\mathrm{rot}}$ was solved by
using the {\it presumably known} $P_{\mathrm{act}}$
\citep[][their Fig. 32]{Jet17}.
Unfortunately,
we do not have the ``luxury'' of knowing 
the $P_{\mathrm{rot}}$ period 
of FK Com {\it before} the analysis.

If the \JHLarg ~is true,
we must conclude that all our earlier one-dimensional
period analyses have given spurious results,
because these results suffer from \Inc.
The same applies to all one-dimensional
period finding analyses of starspots done
by others during the past 70 years
since \citet{Kro47} published his first
observations of starspots.
Due to this \Inc,
neither one of the two $P_{\mathrm{act}}$ and $P_{\mathrm{rot}}$
periods of FK Com can be fixed before the analysis.
These two periods must be solved simultaneously.
At this point,
we have a serious problem,
because our current study of FK Com photometry
has arrived at a dead end: we can not 
use the $t_{\mathrm{min,1}}$ minima epochs 
of Table \ref{electric} to
get an unambiguous
solution for the numerical  
$P_{\mathrm{rot}}$ or $P_{\mathrm{act}}$ period
values
of FK Com
(the \problemfour ~and the \problemfive).

\subsection{Sum of two different frequency 
sinusoids}
\label{explanation}

In Sect. \ref{simulations}, we assumed that the
\twopmodel ~is correct and simulated 
artificial data with
this model. Here, we show {\it how} 
these results could have
been anticipated mathematically,
and especially the \Inc.
The mathematical relations presented in this section
explain all graphical results presented earlier in 
Figs. \ref{figsimulated1} and \ref{figsimulated2}.

If $f_1 \approx f_2$, 
the solution for the sum $s(t)$ of sinusoidal light curves 
$s_1(t)$ and $s_2(t)$ of Eqs. \ref{sone} and \ref{stwo}
depends {\it only} on the amplitudes $a_1$ and $a_2$. 
The solutions are different for
\begin{eqnarray}
a_1 & =    & a_2 \label{fsame}    \\
a_1 & \neq & a_2. \label{fnotsame}
\end{eqnarray}

If Eq. \ref{fsame} is true, then
\begin{eqnarray}
s(t)=s_1(t)+s_2(t)= a_a(t) \sin{(2 \pi f_a t)},
\label{equalfsum}
\end{eqnarray}
where the amplitude is
\begin{eqnarray}
a_a(t)=a_1\sqrt{2+2\cos{[2\pi(f_1-f_2)t]}}
\label{ampsame}
\end{eqnarray}
and the frequency is
\begin{eqnarray}
f_a=(f_1+f_2)/2.
\label{freqsame}
\end{eqnarray}

\noindent
This frequency of the sum $s(t)$
remains constant $f=f_a$.
Hence, the $P_3$ periods detected with 
the \lsmethod ~concentrate between 
$P_1$ and $P_2$ in Fig. \ref{figsimulated1}b.
The amplitude $a_a(t)$ of Eq. \ref{ampsame} varies between 
$a_{\mathrm{a,min}}=0$ and
$a_{\mathrm{a,max}}=2a_1$ 
with the lap cycle period $P_{\mathrm{lap}}$.
This regularity is seen in Fig. \ref{figsimulated1}c.
Finally, an abrupt $f_a/2$ phase shift of $s(t)$  
occurs when $a_a(t)$ goes to zero,  
as seen in Fig. \ref{figsimulated1}d.

If Eq. \ref{fnotsame} is true, we assume that $a_1<a_2$.
In this case, the sum is
\begin{eqnarray}
s(t)=s_1(t)+s_2(t)=a_b(t) \sin{[2 \pi f_2 t + \phi_b(t)] },
\label{unequalfsum}
\end{eqnarray}
where the amplitude is
\begin{eqnarray}
a_b(t)=\sqrt{a_1^2+a_2^2+2 a_1 a_2 \cos{[2 \pi (f_1-f_2)t]}}
\label{ampnotsame}
\end{eqnarray}
and the phase shift is
\begin{eqnarray}
\phi_b(t)=\arctan{ 
\left[
{
{a_1 \sin{[2\pi(f_1-f_2)t]} }
\over
{a_1\cos{[2\pi(f_1-f_2)t]+a_2}}
}
\right]
}.
\end{eqnarray}
The $[\phi_b(t)]$ units are radians.

The amplitude $a_b(t)$ of Eq. \ref{ampnotsame} varies regularly between 
\begin{eqnarray}
a_{\mathrm{b,min}}=a_2-a_1
\label{minamp}
\end{eqnarray}
and
\begin{eqnarray}
a_{\mathrm{b,max}}=a_1+a_2
\label{maxamp}
\end{eqnarray}
during the lap cycle period $P_{\mathrm{lap}}$,
as seen in Fig. \ref{figsimulated2}c.

The phase modulation $\phi_b(t)$
induces frequency changes of $s(t)$ between 
\begin{eqnarray}
f_{\mathrm{b,max}}=f_2+{{a_1(f_1-f_2)}\over{a_2+a_1}}
\label{upperf}
\end{eqnarray}
at the $a_{\mathrm{b,max}}$ epochs, and
\begin{eqnarray}
f_{\mathrm{b,min}}=f_2-{{a_1(f_1-f_2)}\over{a_2+a_1}}
\label{lowerf}
\end{eqnarray}
at the $a_{\mathrm{b,min}}$ epochs. 
These frequency changes of $s(t)$ 
at both sides of $f_2$ are repeated regularly 
during $P_{\mathrm{lap}}$.
The frequency $f_2$ of the stronger $s_2(t)$
sinusoid $(a_1<a_2)$ dominates these frequency 
changes of $s(t)$.
The $P_3$ changes in Fig. \ref{figsimulated2}b 
concentrate
on the level of the stronger $P_2$ periodicity.

The maximum $\phi_b(t)$ phase shift is 
\begin{eqnarray}
\phi_{\mathrm{b,max}}=\arcsin{(a_1/a_2)}
\label{ampratio2}
\end{eqnarray}
to both directions. 
The $[\phi_{\mathrm{b,max}}]$ units 
are radians.
For $a_1=a_2$, the maximum phase shift $\phi_{\mathrm{b,max}}$
is $\pi/2$ radians, or equivalently $(\pi/2)/(2 \pi)=1/4$
in dimensionless phase units with $f_2$.
Since this shift  at $a_{\mathrm{b,min}}$
occurs to both directions,
its full range is
\begin{eqnarray}
\Delta \phi_{\mathrm{b}}= [\arcsin{(a_1/a_2)}]/\pi
\label{abruptphase}
\end{eqnarray}
in dimensionless phase units.
This gives $\Delta \phi_{\mathrm{b}}=0.5$ when $a_1=a_2$.
No phase shift can exceed this limit.
This abrupt phase shift occurs 
also when $a_1 \ne a_2$, but it is smaller.
For example, the $a_1=0.^{\mathrm{m}}048$ and $a_2=0.^{\mathrm{m}}050$ 
combination causes an abrupt $\Delta \phi_{\mathrm{b}}=0.41$ phase shift
at $a_{\mathrm{b,min}}$. 
The predicted phase shift in Fig. \ref{figsimulated2}d 
is only $\Delta \phi_{\mathrm{b}}=0.17$. 

In conclusion, the {\it observed} amplitude,
period and mimimum
of the sum $s(t)$ fluctuate regularly with the lap
period cycle
$P_{\mathrm{lap}}$ when the condition of Eq. \ref{fnotsame} is true. 

\subsection{Indirect approach }
\label{indirect}

Now that we know all these different mathematical relations
for the sum
$s(t)=s_1(t)+s_2(t)$
presented in the previous 
Sect. \ref{explanation},
it is still next to impossible to figure out
what happens in some particular sample of photometry.
These mathematical relations do, however,
give us some hope in resolving the $P_{\mathrm{act}}$
and $P_{\mathrm{rot}}$ dilemma of FK Com,
because they open up several indirect approaches.
Here, indirect means that we do not solve the
$P_1$ and $P_2$ periods simultaneously and directly.
These indirect approaches are 
valid only if the real light curves
$g_1(t)$ and $g_2(t)$
are sinusoids (Eqs. \ref{sone} and \ref{stwo}).
For example, 
the relations of Eqs. \ref{equalfsum} and \ref{unequalfsum}
show that the shape of the sum curve $s(t)$
is sinusoidal with only one minimum and maximum.
If this observed $s(t)$ light curve has two minima,
the indirect approaches presented below
are no longer valid because at least one of
the real light curves
$g_1(t)$ and $g_2(t)$ can not be a sinusoid.

It has already been mentioned 
in Sect. \ref{amplitudeinc}
that the period
analysis of the observed light curve
amplitudes $A_3$ may reveal the lap
cycle period $P_{\mathrm{lap}}$.
The relation of Eq. \ref{lapcycle} is
easier to manipulate in the frequency domain,
$f_{\mathrm{lap}}=|f_1-f_2|=\pm (f_1-f_2)$. This gives
\begin{eqnarray}
f_1 = \pm f_1 = f_{\mathrm{lap}}\pm f_2
\label{smartone}
\end{eqnarray}
We can use the equality $f_1=\pm f_1$,
because a negative $f_1$ value makes no sense
in this particular context.
Assuming that 
$f_2=P_{\mathrm{orb}}^{-1} \approx
P_{\mathrm{rot}}^{-1}$ in CABSs,
the above relation gives two solutions
for $f_1=P_{\mathrm{act}}^{-1}$ when 
$P_{\mathrm{lap}}=f_{\mathrm{lap}}^{-1}$ is known.
If the signal of the 
$P_{\mathrm{orb}}$ period does not dominate the light curves,
the migration of the light curve minimum epochs is tilted 
as a function of phase $\phi_2$ with $P_2=P_{\mathrm{orb}}=f_2^{-1}$ 
(Eq. \ref{phasetwo}).
During these time intervals,
the forwards or backwards migration of 
these minima reveals if
$P_{\mathrm{act}}^{-1}=f_1=f_{\mathrm{lap}}+f_2$ 
or 
$P_{\mathrm{act}}^{-1}=f_1=f_{\mathrm{lap}}-f_2$.
Unfortunately, 
this relation of Eq. \ref{smartone}
does not help with any CASS,
like FK Com.

There are three migration alternatives for 
the observed $s(t)=s_1(t)+s_2(t)$ light curve minima

\begin{description}

\item  \migrone: If $a_1 \approx a_2$ 
(Eqs. \ref{sone} and \ref{stwo}),
the migration is linear 
with $(f_1+f_2)/2$ (Eq. \ref{freqsame})

\item  \migrtwo: If $a_1 < a_2$
(Eqs. \ref{sone} and \ref{stwo}),
the migration is linear  
with $f_2$ and fluctuating
(Eqs. \ref{upperf} and \ref{lowerf})

\item  \migrthree: If $a_1 > a_2$
(Eqs. \ref{sone} and \ref{stwo}),
the migration is linear 
with $f_1$ and fluctuating
(Eqs. \ref{upperf} and \ref{lowerf})

\end{description}

\noindent
It is not easy to unambiguously separate these
three alternatives from each other in the real
photometric observations.
Nevertheless, we know at least that the
stronger signal usually dominates the linear
fluctuating migration (Eqs. \ref{upperf}-\ref{lowerf}), 
because it is reasonable to assume that
the cases $a_1 \approx a_2$ occur
less frequently than the cases $a_1<a_2$ or $a_1>a_2$.
When the observed $s(t)$ 
light curve minima $t_{\mathrm{min,1}}$
are analysed, e.g. with the \kmethod, 
the result can be $f_1$, $f_2$ or $(f_1+f_2)/2$,
or none of these.
Different migration alternatives of CABSs
were displayed in \citet[][Figs. 15-27: 
tilted non-stationary
and horizontal stationary lines in panels ``a'']
{Jet17}.
We know now that there are at least
three different migration alternatives 
even in the most simple case of two
interacting sinusoids $s_1(t)$ and $s_2(t)$.
For example, the minima of II Peg
first concentrated close
to $\phi_{\mathrm{orb}}=0.25$ between 
the years 1988 and 1997,
and then these minima
began to wander regularly away 
from this phase
\citep[][their Fig. 27a]{Jet17}.
All available $t_{\mathrm{min,1}}$ light curve
minima were used to determine the $P_{\mathrm{act}}$
value of this CABS.
However, we know now that
the result for $P_{\mathrm{act}}$ was not correct,
because some $t_{\mathrm{min,1}}$ values,
like those  between 
the years 1988 and 1997,
were certainly
connected $P_{\mathrm{orb}}$,
but not to $P_{\mathrm{act}} \neq P_{\mathrm{orb}}$.
For any particular star,
the \kmethod ~may detect 
$f_1$, $f_2$ or $(f_1+f_2)/2$ 
depending on which one of the
\migrone, \migrtwo ~or \migrthree ~modes dominates.
The long-term photometry of any CABS or CASS
is a mixture of all
these three migration alternatives, 
because the amplitudes $A_1=2a_1$ and $A_2=2a_2$
can change.


It may be possible to identify the
time intervals when one of the above 
mentioned three migration modes
dominates. 
For example, it is usually easy to notice
the {\it particular}
time intervals when $P_{\mathrm{orb}}$ dominates
the $s(t)$ light curve minima of some
CABS \citep[][e.g. II Peg]{Jet17}. 
It may be possible to
solve $P_{\mathrm{act}}$ by analysing
the $t_{\mathrm{min,1}}$ values of the {\it other} 
remaining time
intervals with the \kmethod.
Finally, the observing seasons displaying
a \flip ~event 
represent the \migrone ~case having
an exceptional amplitude and
periodicity combination.
The detection of the 
$P_{\mathrm{orb}}$ or $P_{\mathrm{act}}$ periods
is impossible during these observing seasons
when $A_1 \approx A_2$,
because an infine number of $P_1=f_1^{-1}$ and $P_2=f_2^{-1}$
period combinations can induce the observed
period $P_3=[(f_1+f_2)/2]^{-1}$.

In Sect. \ref{SIgeneral},
we noted that our $|\Delta \Omega|$ values
in Table \ref{falsediff}
may be underestimates.
The $P_{\mathrm{act}}$ values of some stars
published by  \citet{Jet17} are too close to $P_{\mathrm{orb}}$,
because they used {\it all} $t_{\mathrm{t_{min,1}}}$
values in their \kmethod ~analysis.
{\it Had} they excluded the $t_{\mathrm{min,1}}$
values connected to the $P_{\mathrm{orb}}$ periodicity
before their \kmethod ~analysis, 
some $|P_{\mathrm{orb}}-P_{\mathrm{act}}|$ differences
would have been larger. Hence, we 
have underestimated the $|\Delta \Omega|$ values
in our Table \ref{falsediff}.
In some cases, the \kmethod ~may even have 
detected only the $P_{\mathrm{orb}}$ periodicity, because 
$P_{\mathrm{act}}$ never dominated the observed light curve.
This may have happened for example
with DM UMa 
\citep[][Fig. 15a]{Jet17}.

Regardless of the minima-\Inc,
the three migration alternatives
\migrone, \migrtwo ~and \migrthree ~do 
provide the {\it qualitative} answers to 
the \problemfour, 
the \problemfive ~and the \problemsix,
which are given in 
\yhteys{\paperii ~(Sect. 12)}.

There are other 
indirect ways to
obtain less important additional information. 
Firstly,
the ratio $a_1/a_2$ can be estimated 
from the range of $\phi_{\mathrm{min,1}}$ 
fluctuation (Eq. \ref{ampratio2}).
Secondly,
if the observed $s(t)$ light curve amplitude $A_3$
repeats some regular pattern, the
amplitudes $A_1=2a_1$ and $A_2 = 2 a_2$ can 
be solved from Eqs. \ref{minamp}
and \ref{maxamp}. This gives
\begin{eqnarray}
A_1 & = & (A_{\mathrm{3,max}}-A_{\mathrm{3,min}})/2 
\label{estimateamp1} \\
A_2 & = & (A_{\mathrm{3,max}}+A_{\mathrm{3,min}})/2 
\label{estimateamp2}3
\end{eqnarray}
where $A_{\mathrm{3,max}}$ and $A_{\mathrm{3,min}}$
are the maximum and minimum values of the
{\it observed} amplitude $A_3$.

Although the number of possible
$g_1(t)$ and $g_2(t)$ combinations is infinite,
the above indirect approaches 
give us some hope of detecting the 
$P_{\mathrm{act}}$ and $P_{\mathrm{rot}}$ 
periodicities of FK Com.
However, these indirect 
approaches rely on the assumption that
the observed $s(t)$ light curve is a sum of two
sinusoids.
Even if this assumption were true,
the interpretation of the observed light curves
of FK Com
would remain ambiguous, especially due to \Inc.
We have obtained {\it qualitative} answers to 
the \problemfour, the \problemfive ~and the \problemsix,
but the {\it quantitative} estimates for the
numerical values of $P_{\mathrm{act}}$ and  $P_{\mathrm{rot}}$
are still missing.
Hence, we are
still at the dead encountered in Sect. \ref{deadend}.

\section{Discussion}
\label{discussionone}

\citet{Jet17} presented the general light 
curve model for CABSs and CASSs.
This model was based on the \JHLarg.
We define the complex \twopmodel ~$g_{\mathrm{C}}(t)$ and
the simple \onepmodel ~$g_{\mathrm{S}}(t)$ in Sect. \ref{models}
(Eqs. \ref{gtwomodel} and \ref{gonemodel}).

The models of one-dimensional period finding methods,
like the sinusoidal \lsmodel,
have unique phases, and the shape of the light curve model
does not change.
These models can not describe the light curve of
the \twopmodel, because
there are no unique phases 
for presenting the {\it observed} $g_{\mathrm{C}}(t)$
light curve (Eq. \ref{gtwomodel}).
The shape of this $g_{\mathrm{C}}(t)$
light curve is constantly changing.
Therefore, the $g_{\mathrm{C}}(t)$
light curve can be studied only as a function of time, 
not as a function of phase. 
The only unique phases are 
the $\phi_1$ and $\phi_2$
phases of the {\it real} $g_1(t)$ and $g_2(t)$
light curves
(Eqs. \ref{phaseone} and \ref{phasetwo}). 

The complex $g_{\mathrm{C}}(t)$ model and
the simple  $g_{\mathrm{S}}(t)$ model are nested
(Sect. \ref{nested}).
They are the same model when
$f_1=f_2$ (\firstcase),
and when $g_1(t)$ has $A_1=0$ (\secondcase)
or
$g_2(t)$ has $A_2=0$ (\thirdcase).
If none of these three cases occurs,
we can compare the complex $g_{\mathrm{C}}(t)$ model and
the simple $g_{\mathrm{S}}(t)$ model with
the statistical criterion of Eq. \ref{oneortwo},
and choose which one is the better model for the data.
Such comparisons will be done later 
in  \yhteys{\paperii ~(Sects. 2-4)}.

The earlier \twopmodel ~analysis of thirteen CABSs
relied on the fixed
$P_{\mathrm{act}}$ and $P_{\mathrm{rot}}\approx P_{\mathrm{orb}}$
values
\citep{Jet17}.
One season of FK Com photometry was also studied
with the \twopmodel ~by first solving the fixed
$P_{\mathrm{act}}$ value, and then solving the fixed
$P_{\mathrm{rot}}$ value 
\citep[][Sect. 6.2]{Jet17}.

In Sects. \ref{solarsdr}-\ref{stellarsdr},
we draw attention to the ``sobering reminder'' 
presented by 
\citet{How94}, which may mean that
all results obtained for the stellar \sdr ~with 
the \simethod ~and the \lcmethod ~are not
necessarily true,
especially if the solar-stellar connection holds.
Nevertheless, those earlier SDR results certainly
contradict the \JHLarg ~\citep[e.g.][Sect. 7]{Str09}.
For example,
if this argument is true, the earlier \simethod ~studies
should have detected the two constant periods $P_{\mathrm{act}}$
and  $P_{\mathrm{rot}} \approx P_{\mathrm{orb}}$ of the
\JHLarg ~(the \problemone).
We argue that the parameters
$|k|$ (Eq. \ref{si_k})
and 
$|\Delta \Omega|$ (Eq. \ref{si_omega})
of thirteen CABSs in Table \ref{falsediff}
support the idea that
the earlier \simethod ~studies could have already
detected these two periodicities.
We also present {\it general} and {\it particular} 
evidence for this
(Sects. \ref{SIgeneral} and \ref{SIparticular}).
Long-lived starspots rotating with a constant angular velocity
have been detected in many \simethod ~studies
\citep[e.g.][]{Ber98B,Kor00,Str00A}.
These results provide our answer to the \problemone.

The \lcmethod ~gives several estimates for stellar \sdr,
like the $|k|$ or $Z$ parameters 
(Eqs. \ref{k_one}, \ref{k_two} and \ref{zvalue}).
The values of these parameters are computed from 
the observed changes of the photometric rotation period 
under the assumption that $P_{\mathrm{phot}} \approx P_{\mathrm{rot}}$
(Sect. \ref{LCparameters}).
Numerous \lcmethod ~studies have confirmed
that the stellar \sdr ~decreases 
when $P_{\mathrm{phot}}$ decreases
(Sect. \ref{lcmethodresults}).
These $P_{\mathrm{phot}} \approx P_{\mathrm{rot}}$ 
changes in the observed
light curves of CABSs and CASSs
contradict the \JHLarg ~(the \problemtwo).

The $P_{\mathrm{phot}} \approx P_{\mathrm{rot}}$ values have
usually been determined with the one-dimensional period
finding methods, like the \lsmethod.
We simulate artificial data 
with the most simple \twopmodel: the sum of two sinusoids 
(Sect. \ref{simulations}).
These simulated data are analysed with 
the \lsmethod ~(Sect. \ref{test1}-\ref{test2}).
The results for the light curve
periods $(P_3)$,
amplitudes $(A_3)$
and 
minimum epochs $(t_{\mathrm{1,min}})$
reveal an effect that we refer to as ``\Inc''
(Sect. \ref{incomp}).
The observed values of all these three parameters change,
although the underlying \twopmodel ~does not change
at all (Sects. \ref{periodinc}-\ref{minimainc}).
Due to this \Inc ~effect,
two observers {\it observing} 
the light curve of {\it same} star 
during the {\it same} time interval
will get {\it different}
$P_3$, $|k|$ and $Z$ estimates with the \lsmethod,
or with any other one-dimensional
period finding method.
The \Inc ~of the $P_3$ periods detected with the
\lsmethod, or with any other one-dimensional period finding
method, provides answers 
to the \problemtwo ~and the \problemthree.
This \Inc ~effect does not only contaminate
the \lcmethod ~results, because it can also
mislead the identification of starspots in the surface images
obtained with the {\simethod}s (Sect. \ref{mapinc}).

Due to the minima-\Inc ~effect, our study arrives at a dead end
in Sect. \ref{deadend}.
The \kmethod ~analysis of the minimum epochs $t_{\mathrm{min,1}}$ determined
with the one-dimensional \cpsmethod ~does not 
give us an unambiguous $P_{\mathrm{act}}$ value for FK Com.
The relation $P_{\mathrm{rot}} \approx P_{\mathrm{orb}}$ can not
be used, because FK Com is not a binary.
In other words, we have no {\it qualitative} or {\it quantitative}
answers for the \problemfour,
the \problemfive ~or the \problemsix.

The most simple \twopmodel, the sum of two sinusoids,
is already studied in Sect. \ref{simulations}.
We show how 
the periods, the amplitudes and the minimum phases of
this particular \twopmodel ~can be 
solved analytically (Sect. \ref{explanation}).
These solutions allow
us to obtain indirect information 
of the {\it real} light curves (the two sinusoids)
that cause the {\it observed} light curve (the sum of 
two sinusoids).
For example, this indirect approach reveals the three migration
alternatives \migrone, \migrtwo ~and \migrthree ~(Sect. \ref{indirect}).
These migration alternatives provide {\it qualitative}
answers to the \problemfour,
the \problemfive ~and the \problemsix.
For example,
perfect \flip ~events $\Delta \phi_b=0.5$
occur when the two real light curves are equal amplitude
sinusoids (Eq. \ref{abruptphase}).

The above indirect solutions are valid {\it only, if}
the correct complex $g_{\mathrm{C}}(t)$ 
model is a sum of two sinusioids.
In this case, the sum curve  $g_{\mathrm{C}}(t)$ is also
a sinusoid (Eqs. \ref{equalfsum} and \ref{unequalfsum}).
For example,
if the {\it observed} $g_{\mathrm{C}}(t)$
light curve has two minima,
then at least one of the real $g_1(t)$ and $g_2(t)$ light curves
can not be a sinusoid. 
In this case, the analytical relations for a sum of sinusoids
given in  Sect. \ref{explanation}
are no longer valid.
Thus, the {\it quantitative} answers to 
the \problemfour ~and the \problemfive ~are still missing.
The only way out of this dead end
is to formulate a new two-dimensional period analysis
method,
which gives the unambiguous solutions
for all parameters of the \twopmodel.
This is done in \paperii.


\section{Conclusions}
\label{conclusionsone}

In this first paper, we study the real and the simulated
photometry of FK Com to verify
validity of the following argument made by
\citet{Jet17}:
\begin{description}
\item[-] \JHLarg: \JHLtext
\end{description}
\noindent
This argument is not supported by
the current widely held views 
of the stellar surface differential rotation
and the starspots \citep[e.g.][Sect. 7]{Str09}.
Therefore,
we make six specific questions which 
undermine this argument (Sect. \ref{oneintro}).
Our aim is to answer all these questions.

We present general and particular
evidence that the long-lived starspots 
predicted by the \JHLarg ~have
already been detected in 
the earlier \simethod ~studies.
Our \lsmethod ~analysis of the real and the simulated
photometry of FK Com reveals that the results obtained
with {\it any} one-dimensional
period analysis method suffer from \Inc.  
{\it If} the \JHLarg ~is true,
the one-dimensional period finding methods,
like the \lsmethod, detect 
spurious period, amplitude and minimum epoch estimates
from the observed light curves. 
These spurious estimates have nothing to do with
the real periods, the real amplitudes and the real
minimum epochs of the two
real light curves of the \JHLarg.
This would mean that all earlier one-dimensional period 
analysis studies of the light curves of
chromospherically active stars have given spurious results
since the starspots were discovered by \citet{Kro47}.
The spurious period changes detected with these {\lcmethod}s
would explain why the light curves and the surface images
give different surface \sdr ~estimates even for the
same star. We also show that it is possible to obtain 
indirect information of the two {\it real} light light curves,
if the {\it observed} light curve is a sum of
two constant period sinusoids.

In conclusion,
{\it if} the \JHLarg ~is true, we can already answer 
the \problemone, the \problemtwo, 
the \problemthree ~and the \problemsix.
We can also give {\it qualitative} answers to
the \problemfour ~and the \problemfive.
Even if the \JHLarg ~were true,
the {\it quantitative} answers to the
\problemfour ~and the \problemfive
~are still missing, because we can not
solve the unambiguous values for the
$P_{\mathrm{act}}$ and $P_{\mathrm{rot}}$ periods
of FK Com.
Hence, our study arrives at a dead end,
because we can not prove
that the long-lived starspots predicted by
the \JHLarg ~do exist in FK Com.
We solve this problem in our second paper,
where we formulate a new two-dimensional
period finding method, 
solve the real light curves of FK Com and
determine its unambiguous $P_{\mathrm{act}}$ and $P_{\mathrm{rot}}$
period values.
The compact answers to all the above mentioned
six specific questions are given in
\yhteys{\paperii ~(Sect. 12).}

\section*{Acknowledgements}
We thank Thomas Hackman and Ilana Hiilesmaa
for their comments of this manuscript. 
This work has made use of NASA's 
Astrophysics Data System (ADS) services.

\bibliographystyle{mnras}
\bibliography{jetsufk}

\appendix

\section{Abbreviations}

We use the following abbreviations in \paperi ~and \paperii

\begin{description}
\item[-] \kmethod: Kuiper method                     (\paperi: Sect.~\ref{oneintro})
\item[-] CABS: Chromospherically Active Binary Star  (\paperi: Sect.~\ref{oneintro}) 
\item[-] MLC: Mean Light Curve                       (\paperi: Sect.~\ref{oneintro})
\item[-] CASS: Chromospherically Active Single Star  (\paperi: Sect.~\ref{oneintro}) 
\item[-] \sdr: Surface Differential Rotation         (\paperi: Sect.~\ref{oneintro})
\item[-] \JHLarg: \citet{Jet17} argument             (\paperi: Sect.~\ref{oneintro})
\item[-] \simethod: Surface Images measure \sdr     ~(\paperi: Sect.~\ref{oneintro})  
\item[-] \lcmethod: Light Curves measure \sdr        ~(\paperi: Sect.~\ref{oneintro})  
\item[-] TEL=1: T7 telescope photometry              (\paperi: Sect.~\ref{data}) 
\item[-] TEL=2: Ph10 telescope photometry            (\paperi: Sect.~\ref{data}) 
\item[-] SEG: Data segment number                    (\paperi: Sect.~\ref{data}) 
\item[-] \cpsmethod: Continuous Period Search method (\paperi: Sect.~\ref{data})  
\item[-] \cpsmodel: Continuous Period Search model   (\paperi: Sect.~\ref{data})   
\item[-] \Data{x}{y}: TEL=x data of SEG=y            (\paperi: Sect.~\ref{models})
\item[-] \twopmodel: Two Period model                (\paperi: Sect.~\ref{secttwop}) 
\item[-] \onepmodel: One Period model                (\paperi: Sect.~\ref{sectonep}) 
\item[-] \lsmethod: Lomb-Scargle method              (\paperi: Sect.~\ref{lcmethodsect}) 
\item[-] \lsmodel: Lomb-Scargle model                (\paperi: Sect.~\ref{lcmethodsect})
\item[-] \testone: \twopmodel ~test $A_1=A_2$         (\paperi: Sect.~\ref{test1}) 
\item[-] \testtwo: \twopmodel ~test $A_1<A_2$         (\paperi: Sect.~\ref{test2}) 
\item[-] \twopmethod: Two period method              (\paperii: \yhteys{Sect.~2.1})
\item[-] \onepmethod: One period method              (\paperii: \yhteys{Sect.~2.2}) 
\item[-] \wkmethod: Weighted Kuiper method           (\paperii: \yhteys{Sect.~7}) 
\end{description}

\noindent
We introduce these abbreviations in the sections given above
in the parenthesis.
For easy readability,
we do not use these abbreviations 
in our six specific questions
or in our answers to these six questions,
except for the \JHLarg ~abbreviation.

\bsp    
\label{lastpage}
\end{document}